\documentclass[linenumbers]{aastex631}

\usepackage{amsmath}
\usepackage{amssymb}
\usepackage{graphicx}
\usepackage{bm}
\usepackage{mathrsfs}
\usepackage{color}

\usepackage{subfigure}
\usepackage{epstopdf, epsfig}
\usepackage{array} 
\usepackage{stackengine}

\usepackage{dcolumn}
\usepackage{soul}
\stackMath
\usepackage{natbib}

\usepackage{booktabs}
\usepackage{multirow}

\newcommand{\beq}{\begin{equation}}
\newcommand{\eeq}{\end{equation}}

\newcommand{\ben}{\begin{eqnarray}}
\newcommand{\een}{\end{eqnarray}}

\begin{document}

\nolinenumbers

\title{Electron-only reconnection and ion heating in 3D3V hybrid-Vlasov plasma turbulence}

\author{C. Granier}
\affiliation{Max Planck Institute for Plasma Physics, 
            Boltzmannstraße 2, 
            85748 Garching, 
            Germany}

\author{S. S. Cerri}
\affiliation{Université Côte d'Azur, Observatoire de la Côte d'Azur,  CNRS, Laboratoire J.L. Lagrange, \\
            Boulevard de l'Observatoire, CS  34229, 
            06304 Nice Cedex 4}

\author{F. Jenko}
\affiliation{Max Planck Institute for Plasma Physics, 
            Boltzmannstraße 2, 
            85748 Garching, 
            Germany}

\begin{abstract}
\nolinenumbers

We perform 3D3V hybrid-Vlasov simulations of turbulence with quasi-isotropic, compressible injection near ion scales to mimic the Earth's magnetosheath plasma, and investigate the novel electron-only reconnection, recently observed by the NASA's MMS mission, and its impact on ion heating. Retaining electron inertia in the generalized Ohm’s law enables collisionless magnetic reconnection. Spectral analysis shows a shift from kinetic Alfvén waves (KAW) to inertial kinetic Alfvén (IKAW) and inertial whistler waves (IWW) near electron scales. To distinguish the roles of inertial scale and gyroradius ($d_{\rm{i}}$ and $\rho_{\rm{i}}$), three ion beta ($\beta_{\rm{i}} = 0.25, 1, 4$) values are studied. Ion-electron decoupling increases with $\beta_{\rm{i}}$, as ions become less mobile when the injection scale is closer to $\rho_{\rm{i}}$ than $d_{\rm{i}}$, highlighting the role of $\rho_{\rm{i}}$ in achieving an electron magnetohydrodynamic (EMHD) regime at sub-ion scales. This regime promotes electron-only reconnection in turbulence with small-scale injection at $\beta_{\rm{i}} \gtrsim 1$. We observe significant ion heating even at large $\beta_{\rm{i}}$, with $Q_{\rm{i}}/\epsilon \approx 69\%, 91\%, 96\%$ at $\beta_{\rm{i}} = 0.25, 1, 4$ respectively. While ion heating is anisotropic at $\beta_{\rm{i}} \leq 1$ ($T_{\rm i,\perp} > T_{\rm i,\parallel}$), it is marginally anisotropic at $\beta_{\rm{i}} > 1$ ($T_{\rm i,\perp} \gtrsim T_{\rm i,\parallel}$). Our results show ion turbulent heating in collisionless plasmas is sensitive to the separation between injection scales ($\lambda_{\text{inj}}$) and $\rho_{\rm{i}}$, $\beta_{\rm{i}}$, and finite-$k_{\parallel}$ effects, necessitating further investigation for accurate modeling. These findings have implications for other collisionless astrophysical environments, like high-$\beta$ plasmas in intracluster medium, where processes such as micro-instabilities or shocks may inject energy near ion-kinetic scales.

\end{abstract}


\section{Introduction}

Turbulence and magnetic reconnection are fundamental and ubiquitous processes in space and astrophysical plasmas. The former naturally arises from phenomena that release free energy in the system such as, for instance, large-scale shear flows, shocks, and other plasma instabilities triggered by supernovae explosions, jets, accretion processes, and winds~\citep[e.g.,][]{QuataertGruzinovAPJ1999,SchekochihinCowleyPOP2006,BrandenburgLazarianSSRv2013,BrunoCarboneLRSP2013}. This fluctuations' energy is usually injected at large (``fluid'') scales and then is nonlinearly transferred towards smaller and smaller scales, until dissipation into heat and non-thermal particles is achieved at the characteristic microscopic (``kinetic'') scales of the plasma (i.e., the Larmor radius and/or the inertial length of the particle species). Magnetic reconnection, on the other hand, is a micro-scale process that changes the magnetic-field connectivity of energetically unfavorable configurations by releasing excess magnetic energy (e.g., into bulk flows, waves, and non-thermal particles), and can affect a wide range of scales~\citep[e.g.,][]{ZweibelYamadaARAA2009,PucciJPP2020}. Reconnection is indeed an intrinsic element of plasma turbulence, as the latter naturally develops tearing-unstable current sheets on a wide range of scales along its cascade, and current-sheet reconnection can convert magnetic energy into fluctuations and structures that feed back into the turbulent cascade~\citep[e.g.,][]{CarbonePOF1990,HuangBhattacharjeeAPJ2016,CerriCalifanoNJP2017,FranciAPJL2017,LoureiroBoldyrevAPJ2017,MalletJPP2017,PucciAPJ2017,ComissoAPJ2018,DongPRL2018,DongSciAdv2022,PapiniAPJ2019,BorgognoAPJ2022,CerriAPJ2022}.

Traditionally, in collisionless magnetic reconnection, a multiscale structure emerges, featuring a small electron diffusion region (EDR) within a larger ion diffusion region (IDR). In the diffusion regions, the frozen-in condition, which implies a strong coupling between the magnetic field and the plasma, is broken due to non-ideal effects. In the IDR, only ions demagnetize due to the Hall effect, while electrons remain frozen. Ion-scale bidirectional jets pointing away from the reconnection point (called ``X-point'') indeed emerge from this region~\citep{Bir01, Hes01}. Recent studies, however, have unveiled a novel and intriguing phenomenon known as "electron-only reconnection," revealing a unique form of reconnection, in the absence of ion jets.
Electron-only reconnection refers to reconnection events where the current density is dominantly carried by electrons alone. This distinctive form of reconnection has been observed in the turbulent Earth's magnetosheath~\citep{Pha18, StawarzAPJL2019, Sta22}, at the magnetopause~\citep{Hua21}, and in the magnetotail~\citep{Wan20}. In particular, the high-resolution measurements from NASA's Magnetospheric Multiscale (MMS) mission have provided a unique opportunity to investigate electron scale reconnection in greater detail, revealing new features that expand the traditional understanding of reconnection in collisionless plasmas.

Numerical and theoretical studies have played an important role in understanding this phenomenon. Some 2D Particle-in-Cell (PIC) simulations have suggested that when the current sheet's length is less than approximately $10 d_{\rm i}$, ions fail to couple to the newly reconnected field lines, leading to the absence of ion jets~\citep{Pya19}. Additionally, 2D simulations of decaying turbulence, in which the initial injection scale of fluctuations was varied, showed that larger injection scales result in ion-coupled reconnection, while shorter injection scales favor electron-only reconnection~\citep{CalifanoFRP2020, ArroAA2020, Fra22}. 
This finding aligns with in-situ observations by MMS,  which have further reinforced this notion. In the observations reported in \cite{Sta22}, indeed, when the correlation length of turbulence is less than $20 d_{\rm i}$, there is a trend toward thinner current sheets and the possibility of faster electron jets. Additionally, \cite{Bet23} study draws comparisons between the Electron Magnetohydrodynamics (EMHD) regime and the defining features of the electron-only reconnection. The study reveals that certain aspects of such reconnection regime can be accurately described by EMHD equations, which is valid on scales $\ell/d_{\rm i} < 1$. 
The role of the ion gyroradius scale has only been recently investigated in the 2D particle-in-cell simulations conducted by   \cite{Gua23}. They observed that as the ion gyroradius $\rho_{\rm i}$ increases, ion response weakens, enhancing reconnection rates. Notably, electron-only reconnection occurs when $\rho_{\rm i}$  matches the simulation domain size in both strong and weak guide field scenarios.
In our study, we aim to investigate the role of $\rho_{\rm i}$ by covering cases where $\rho_{\rm i} < d_{\rm i}$ and $\rho_{\rm i} > d_{\rm i}$.

It is worth noting that the majority of studies on electron-only reconnection have been conducted in a 2D framework, potentially overlooking several crucial properties compared to the more realistic three-dimensional (3D) scenarios~\citep{How15}. In this work, we employ the hybrid-Vlasov-Maxwell (HVM) code~\citep{ValentiniJCP2007} to perform 3D simulations of plasma turbulence with different ion plasma beta, $\beta_{\rm{i}}=0.25$, $1$, $4$ ($\beta_{\rm{i}}$ is the ratio between the ion thermal pressure and the magnetic pressure) and finite electron-inertia effects (with a reduced ion-to-electron mass ratio $m_{\rm{i}}/m{\rm e}=100$). Adopting a generalized Ohm's law with electron inertia terms indeed enables to capture the EMHD regime at electron scales and to provide a physical mechanism to drive collisionless magnetic reconnection~\citep[e.g.,][]{CalifanoFRP2020}. Moreover, by varying the ion beta, these simulations allow to disentangle the possible different role of the ion inertial length $d_{\rm{i}}$ and ion Larmor radius $\rho_{\rm{i}}$, and so to further explore the most-favorable regime in which electron-only reconnection can occur in 3D kinetic turbulence. 
Despite the much greater computational cost in comparison to a particle-in-cell (PIC) formulation, the grid-based Vlasov approach offers the distinct advantage of significantly reducing numerical noise at the smallest scales of the simulation (which are the relevant ones to study the electron-only reconnection regime).

The remainder of the paper is organized as follows. In Section~\ref{sec:data_set}, we describe the hybrid-kinetic model that is employed in this work~(\S\ref{subsec:HVM_model}) and the setup of our numerical simulations~(\S\ref{subsec:sim_setup}), detailing the motivation of our choice~(\S\ref{subsec:setup_justification}) and explicitly mention also the intrinsic limitations of this study~(\S\ref{subsec:limitations}).
The simulations' results are analyzed in Section~\ref{sec:results}. 
After characterizing the global properties of turbulent fluctuations~(\S\ref{subsec:spectra}), we turn our attention to magnetic reconnection in 3D~(\S\ref{subsec:3Drec_events}) and to the distinctive signatures of electron-only reconnection events~(\S\ref{subsec:e-rec_events}).
Finally, we focus on analysing the turbulent heating of the ions and the inferred ion-to-electron heating ratio~(\S\ref{subsec:ion_heating}).
In Section~\ref{sec:conclusions}, we summarize the results and discuss their implications.

\section{Numerical Data Set}\label{sec:data_set}

\subsection{Hybrid-Vlasov Model}\label{subsec:HVM_model}

In this work we employ the so-called hybrid-Vlasov--Maxwell (HVM) code~\citep{ValentiniJCP2007}.
The hybrid-kinetic approximation assumes a quasi-neutral plasma ($n_{\rm i} = n_{\rm e} \equiv n$) where fully kinetic ions are coupled to an electron fluid through a generalized Ohm's law~\citep[e.g.,][]{WinskeSSRv1985}. 
The quasi-neutrality assumption neglects charge separation, thus constraining this model to phenomena characterized by frequencies lower than the electron plasma frequency.

The evolution of the ion distribution function, denoted as $f(\boldsymbol{x}, \boldsymbol{v}, t)$, follows the Vlasov equation:
\beq
\partial_t f+\boldsymbol{v} \cdot \boldsymbol{\nabla} f+\frac{e}{m_{\mathrm{i}}}\left(\boldsymbol{E}+\frac{\boldsymbol{v}}{c} \times \boldsymbol{B}\right) \cdot \boldsymbol{\nabla}_{\boldsymbol{v}} f=0,
\eeq
while the magnetic field, $\boldsymbol{B}$, evolves accordingly to Faraday's law of induction,
\beq \label{eq:Faraday}
\partial_t \boldsymbol{B}= - c\boldsymbol{\nabla} \times \boldsymbol{E} .
\eeq
The ions are coupled to the electron fluid through a generalized Ohm's law that provides the electric field $\boldsymbol{E}$,
\beq
\left(1-d_{\mathrm{e}}^2 \nabla^2\right) \boldsymbol{E}=-\frac{\boldsymbol{u}_{\mathrm{i}}}{c}\times \boldsymbol{B} + \frac{\boldsymbol{J} \times \boldsymbol{B}}{e n c} +\frac{\boldsymbol{\nabla} P_{\mathrm{e}}}{e n}+\frac{4\pi d_{\mathrm{e}}^2}{c^2} \boldsymbol{\nabla} \cdot\left(\boldsymbol{u}_{\mathrm{i}} \boldsymbol{J} + \boldsymbol{J} \boldsymbol{u}_{\mathrm{i}} - \frac{\boldsymbol{J} \boldsymbol{J}}{en}  \right), \label{ohms}
\eeq
where $d{\rm e}$ represents the electron inertial length, $n$ is the number density, and $\boldsymbol{u}_i$ and $\boldsymbol{u}_e \equiv \boldsymbol{u}_i - \boldsymbol{J} /n$ denote the proton and electron fluid velocities, respectively. Ion's fluid quantities are computed as velocity space moments of $f$, i.e., $n=\int{\rm d}^3\boldsymbol{v} f$ and $\boldsymbol{u}_{\rm i}=n^{-1} \int{\rm d}^3\boldsymbol{v}\, \boldsymbol{v} f$. The current density $\boldsymbol{J}$ is derived from the magnetic field using Ampère's law in the non-relativstic limit $\omega/kc\ll1$ (i.e., without displacement current), $\boldsymbol{J} = \frac{c}{4\pi}\nabla \times \boldsymbol{B}$. 
Finally, for the purpose of this work, we adopt a simple isothermal closure for the electron pressure, namely $P_{\rm e} = n T_{\rm 0e}$. The effect of different electron closures \citep[see, e.g.,][]{FinelliAA2021}, will be investigated in the future.

The generalized Ohm's law (\ref{ohms}) includes electron-inertia effects, represented by the terms proportional to $d_{\rm e}^2$. Electron inertia serves as a physical mechanism capable of driving magnetic reconnection in the collisionless regime, as it can break the frozen in condition. 
Moreover, a finite electron mass introduces the characteristic length scale $d_{\rm e}$ which marks a change in the plasma dynamics and can thus produce another spectral break at $k_\perp d_{\rm e} \sim 1$ in the spectra of collisionless turbulent fluctuations. 
The $d_{\rm e}^2 \nabla^2 \boldsymbol{E}$ term on the left-hand side of (\ref{ohms}) comes from the time derivative of the current in the generalized Ohm's law by using Amp\`ere's and Faraday's laws, and eventually neglecting $\boldsymbol{\nabla} \cdot \boldsymbol{E}$ due to quasi-neutrality, i.e., $\partial \boldsymbol{J} / \partial t = \frac{c}{4\pi}\boldsymbol{\nabla} \times (\partial \boldsymbol{B} / \partial t) = - \frac{c^2}{4\pi} \boldsymbol{\nabla} \times \boldsymbol{\nabla} \times \boldsymbol{E}\approx \frac{c^2}{4\pi}\nabla^2 \boldsymbol{E}$.  
We mention that this approach differs from that employed in \cite{Mun23}, where electron inertia is directly integrated into the generalized Ohm's law without this approximation.

\subsection{Simulations setup}\label{subsec:sim_setup}

The initial background plasma configuration consists of a stationary, spatially homogeneous Maxwellian ion-electron distribution corresponding to a uniform plasma density $n_0=1$, and a reduced mass ratio $m_{\text{i}} / m_{\text{e}} = 100$. 
This plasma in embedded in a uniform magnetic field $\boldsymbol{B}_0= B_0\hat{\boldsymbol{z}}$, with $B_0 = 1$ (in the following parallel ($\|$) and perpendicular ($\perp$) directions will be defined with respect to this background field, unless specified otherwise).
We consider three different initial ion plasma beta values, namely  $\beta_{\rm i} = 0.25$, 
$\beta_{\rm i} = 1$, 
and $\beta_{\rm i} = 4$. 
The temperature ratio $\tau = T_{\rm0i} / T_{\rm0e}$ is chosen to be $\tau= 2.5$ for $\beta_{\rm i} = 0.25$, 
$\tau = 10$ for $\beta_{\rm i} = 1$, 
and $\tau=40$ for $\beta_{\rm i} = 4$. 
These values of $\tau$ correspond to $\beta_{\rm e}=0.1$, which ensures that the electron Larmor radius $\rho_{\rm e}$ is sufficiently smaller than $d_{\rm e}$, thus justifying the choice to neglect finite-Larmor-radius (FLR) corrections while including finite-inertia effects in our electron model.
In fact, the scale at which $k\rho_{\rm e}\sim1$ would correspond to $kd_{\rm i}\sim31$, which is within the range of scale that is affected by numerical dissipation (see below and, e.g., Fig~\ref{fig:Spectra_EB}).
The simulation box is a cube of size $L = 6 \pi d_{\rm i}$ discretized by means of $256^3$ grid points, corresponding to a wavenumber range $0.3\lesssim k d_{\rm i}\lesssim 43$. Numerical filters are employed to remove fluctuation energy only at the smallest scales of the simulation~\citep{LeleJCP1992}.
In velocity space, we cover the range $\left[-7v_{\text{th,i}}, 7v_{\text{th,i}}\right]$ for the case $\beta_{\rm i} = 0.25$ and the range $\left[-5v_{\text{th,i}}, 5v_{\text{th,i}}\right]$ for the $\beta_{\rm i} = 1$ and $\beta_{\rm i} = 4$ cases, and we employ a uniformly distributed grid of $51^3$ points for the $\beta_{\rm i} = 1$ case and $57^3$ points for the $\beta_{\rm i} = 0.25$ and $\beta_{\rm i} = 4$ cases.
The above background is perturbed by adding isotropic $3$D magnetic fluctuations $\delta \boldsymbol{B}=\delta\boldsymbol{B}_\perp + \delta B_\|\hat{z}$ in the wavenumber range $0.3 \lesssim k d_{\rm i} \lesssim 1$ and with  root-mean-square (rms) amplitude $\delta B_{\text{rms},0} / B_0 \simeq 0.5$ (magnetic perturbations are initialized through a vector potential $\delta\boldsymbol{A}$, i.e., $\delta\boldsymbol{B}=\boldsymbol{\nabla}\times\delta\boldsymbol{A}$, thus ensuring the solenoidal character of the magnetic field, $\boldsymbol{\nabla}\cdot\delta\boldsymbol{B}=0$).

\subsection{Connection of our setup with observations and previous turbulent simulations of electron-only reconnection}\label{subsec:setup_justification}

Injecting both parallel and perpendicular magnetic fluctuations with non-negligible amplitude ($\delta B/B\sim0.5$), and doing so up to scales near the ion inertial length ($19\,d_{\rm i}<\lambda_{\rm inj} < 6\,d_{\rm i}$), aims at qualitatively reproducing the typical conditions past the Earth's bow shock, in the magnetosheath. In this environment, turbulence is indeed significantly compressible and characterized by fluctuations with large amplitudes and short correlation lengths~\citep[i.e., $0.5\lesssim\delta B/B\lesssim 1.5$ and $\lambda_{\rm c}\sim 10\,d_{\rm i}$, respectively; see, e.g.,][]{ChenBoldyrevAPJ2017,StawarzAPJL2019}. This is also consistent with the results of previous 2.5D turbulence studies which found that the transition from standard to electron-only reconnection occurs as fluctuations are injected up to scales close to the ion inertial scale~\citep[namely, $\lambda_{\rm inj}\sim10\,d_{\rm i}$;][]{CalifanoFRP2020}.
The choice of the  $\beta_{\rm e}<\beta_{\rm i}$ regime is also meant to reproduce the typical conditions found in the Earth's magnetosheath~\citep{ChenBoldyrevAPJ2017}.
In this work, we extend the above-mentioned previous turbulence studies on electron-only reconnection~\citep[e.g.,][]{CalifanoFRP2020,VegaAPJL2020} from the 2.5D to the 3D geometry. Moreover, we explore different values for the ion beta ({\em viz.}, $\beta_{\rm i}=0.25$, $1$, $4$), in order to possibly disentangle the different role of the two ion-kinetic scales, $\rho_{\rm i}$ and $d_{\rm i}$, compared to the injection lengthscale $\lambda_{\rm inj}$ in the development of (i) the electron-MHD dynamics in 3D turbulence with (moderate) guide field, and of (ii) the consequent occurrence of electron-only reconnection events.

\begin{figure*}[t]
\centering
  \includegraphics[width=0.325\textwidth]{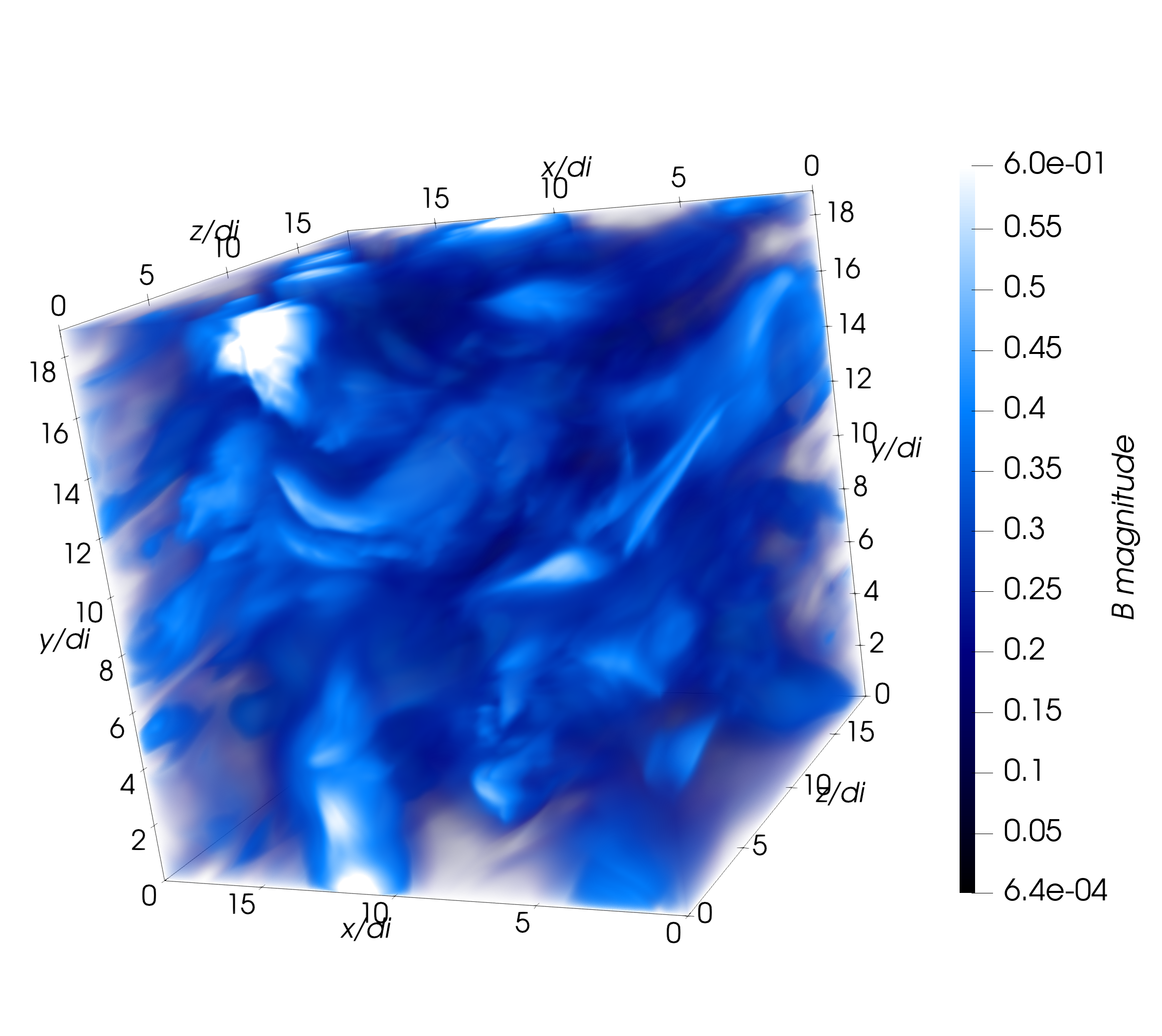} 
  \includegraphics[width=0.325\textwidth]{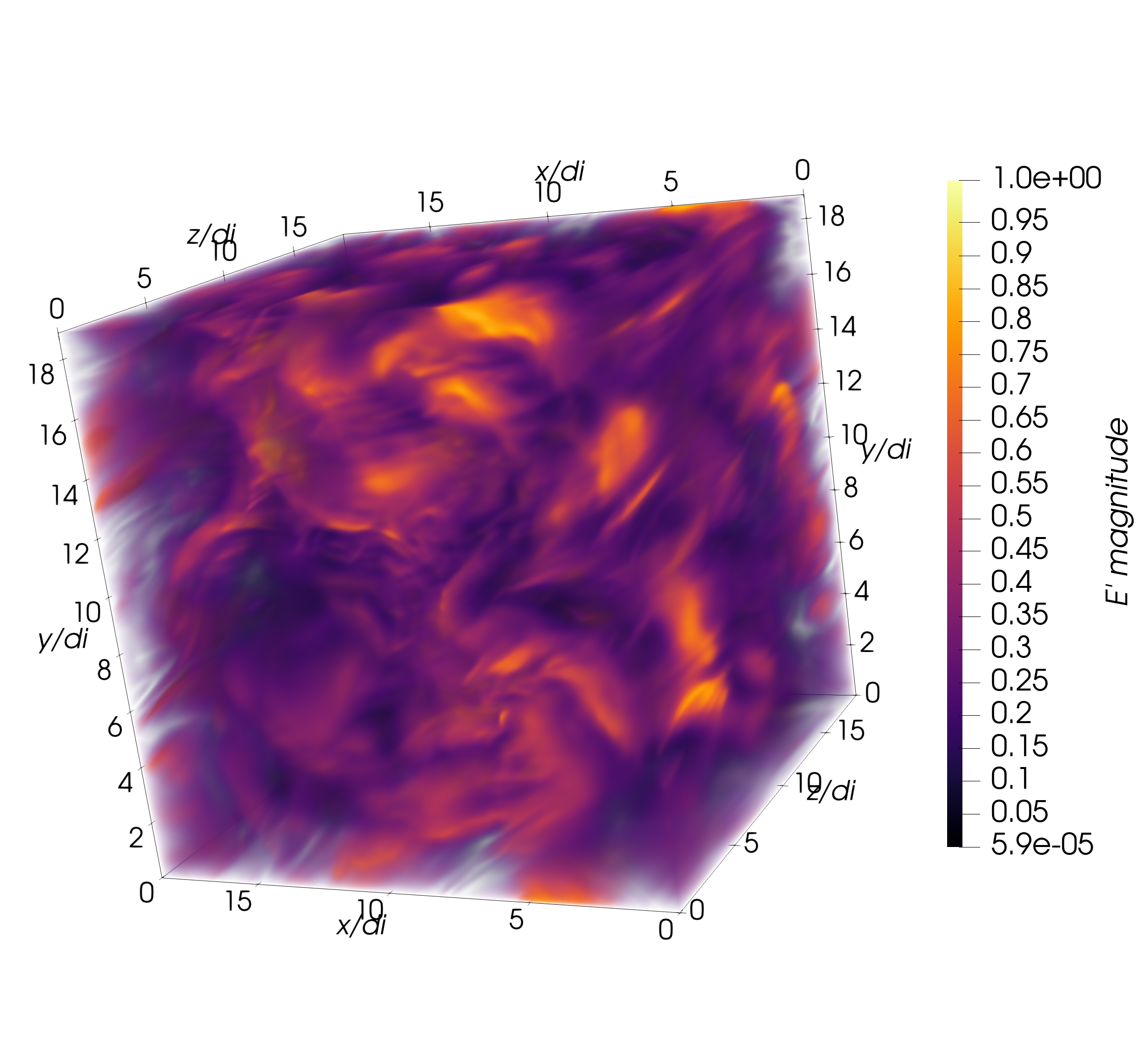} 
  \includegraphics[width=0.325\textwidth]{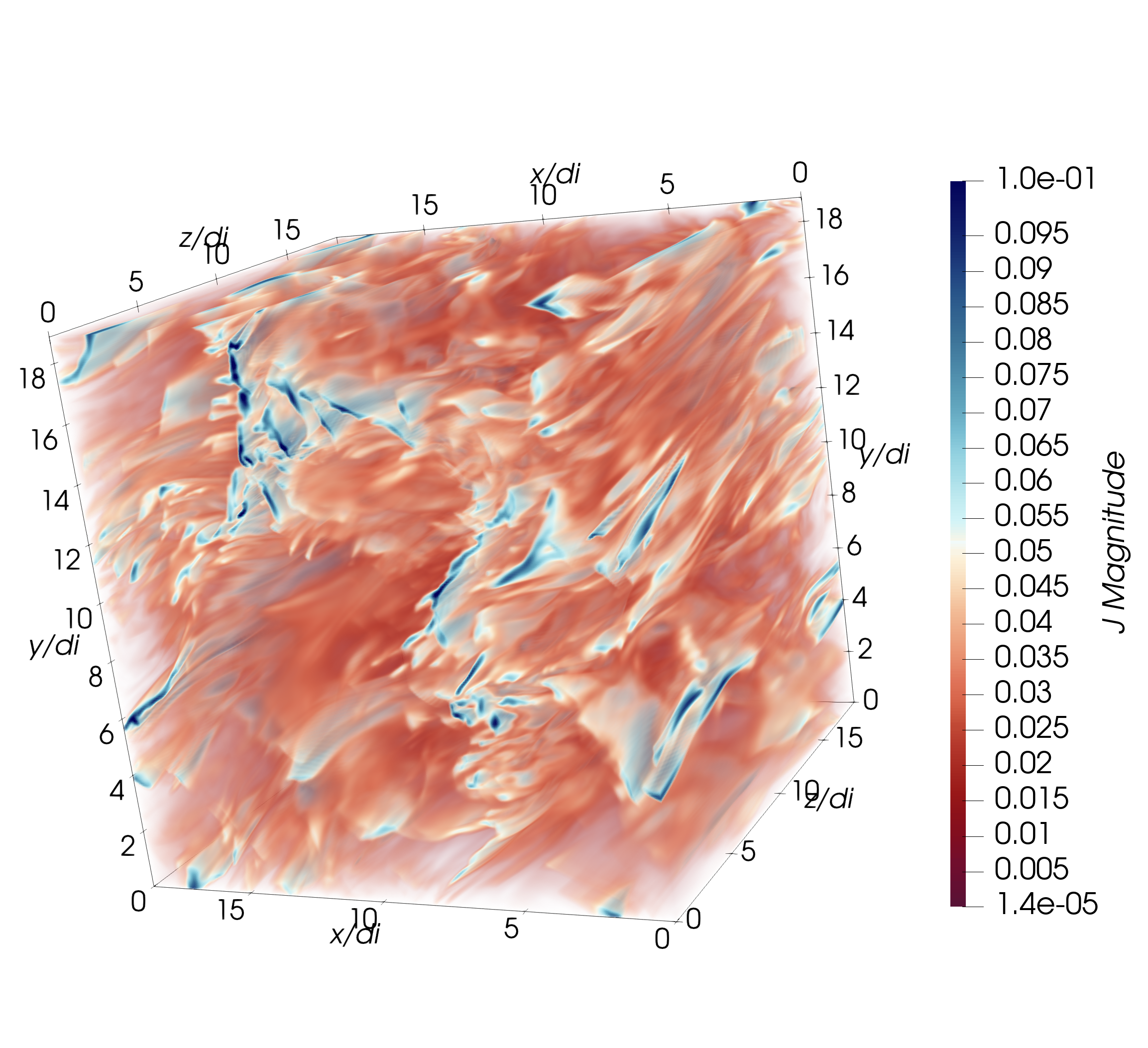} 
  \includegraphics[width=0.325\textwidth]{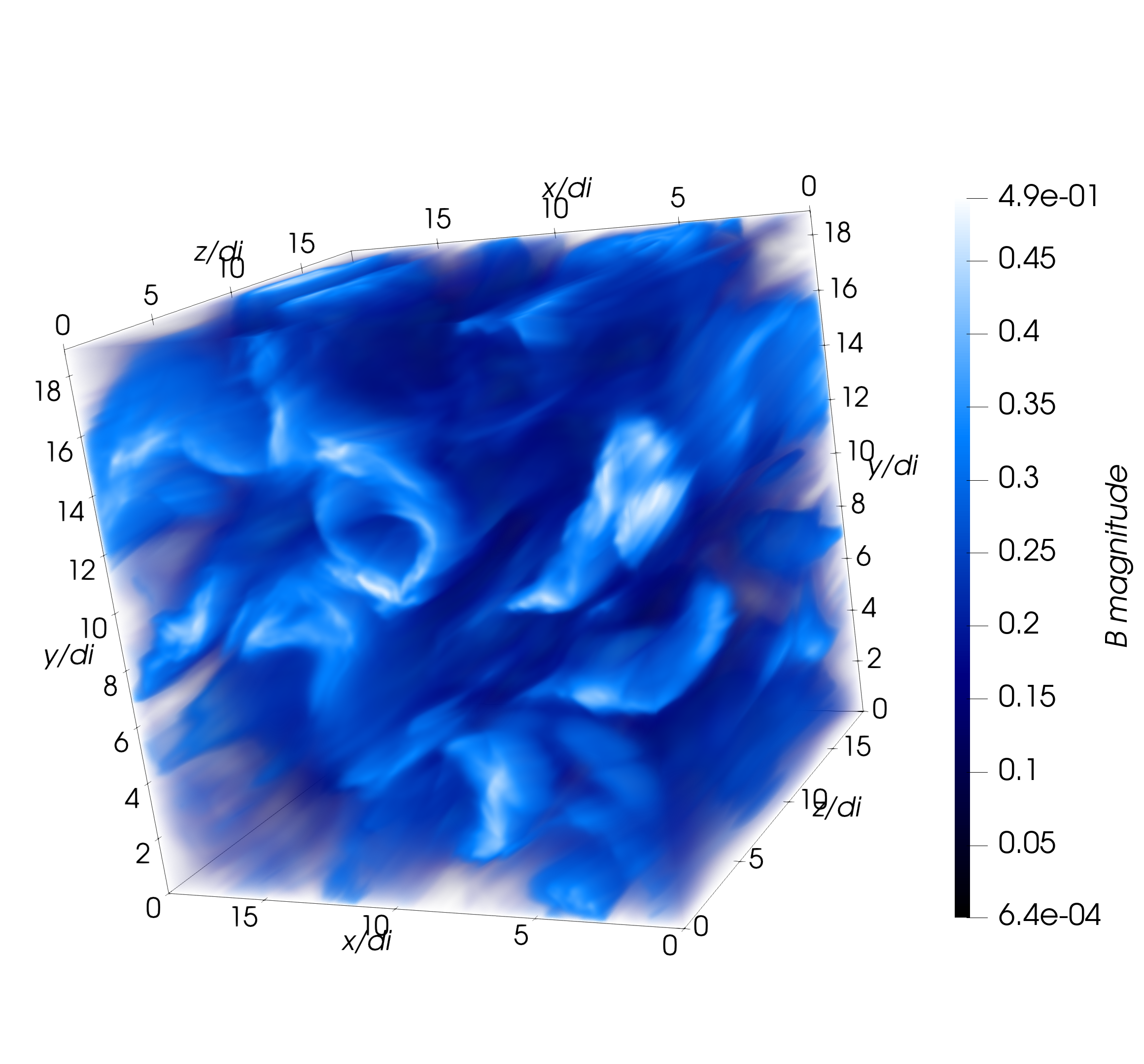} 
  \includegraphics[width=0.325\textwidth]{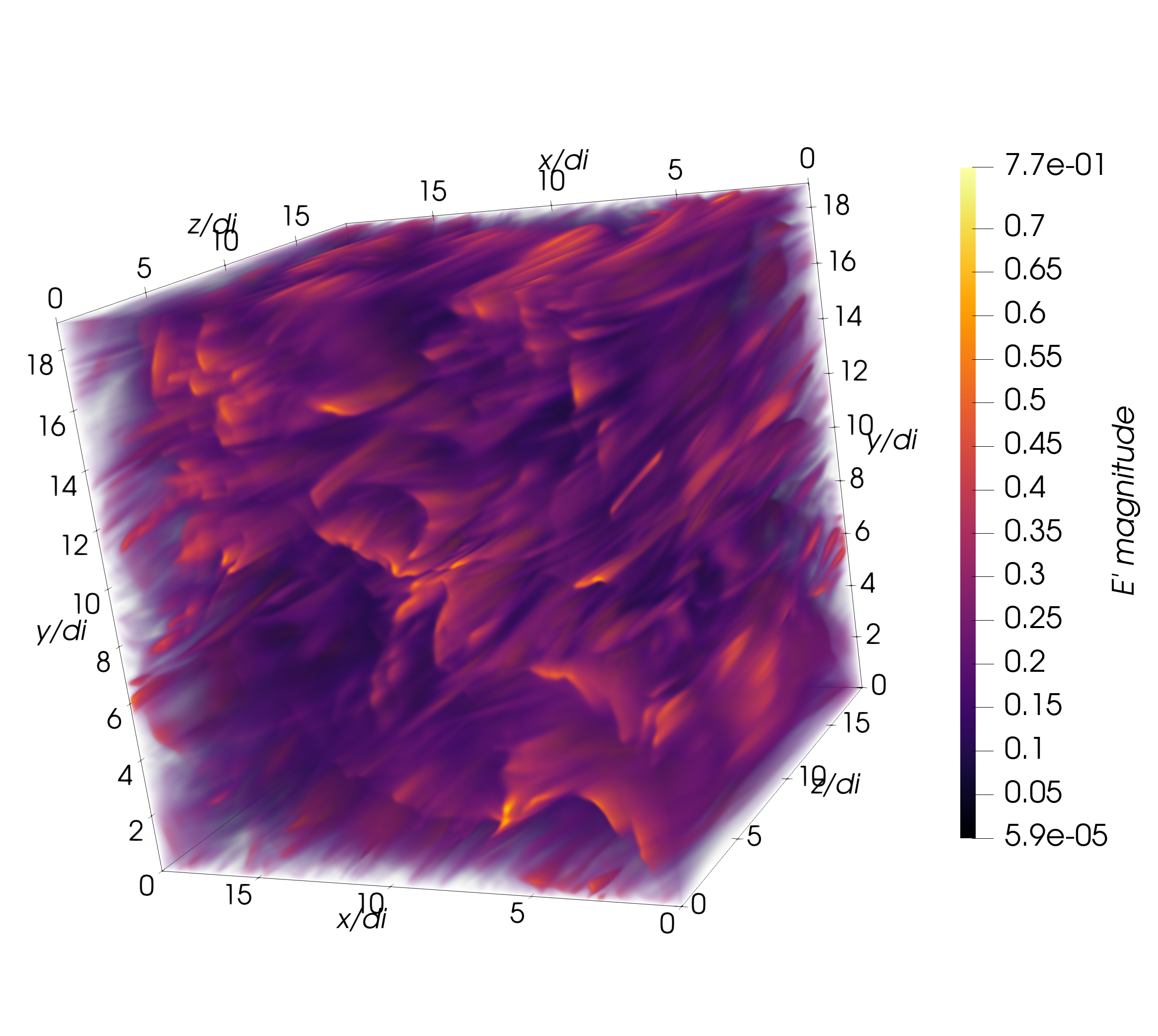} 
  \includegraphics[width=0.325\textwidth]{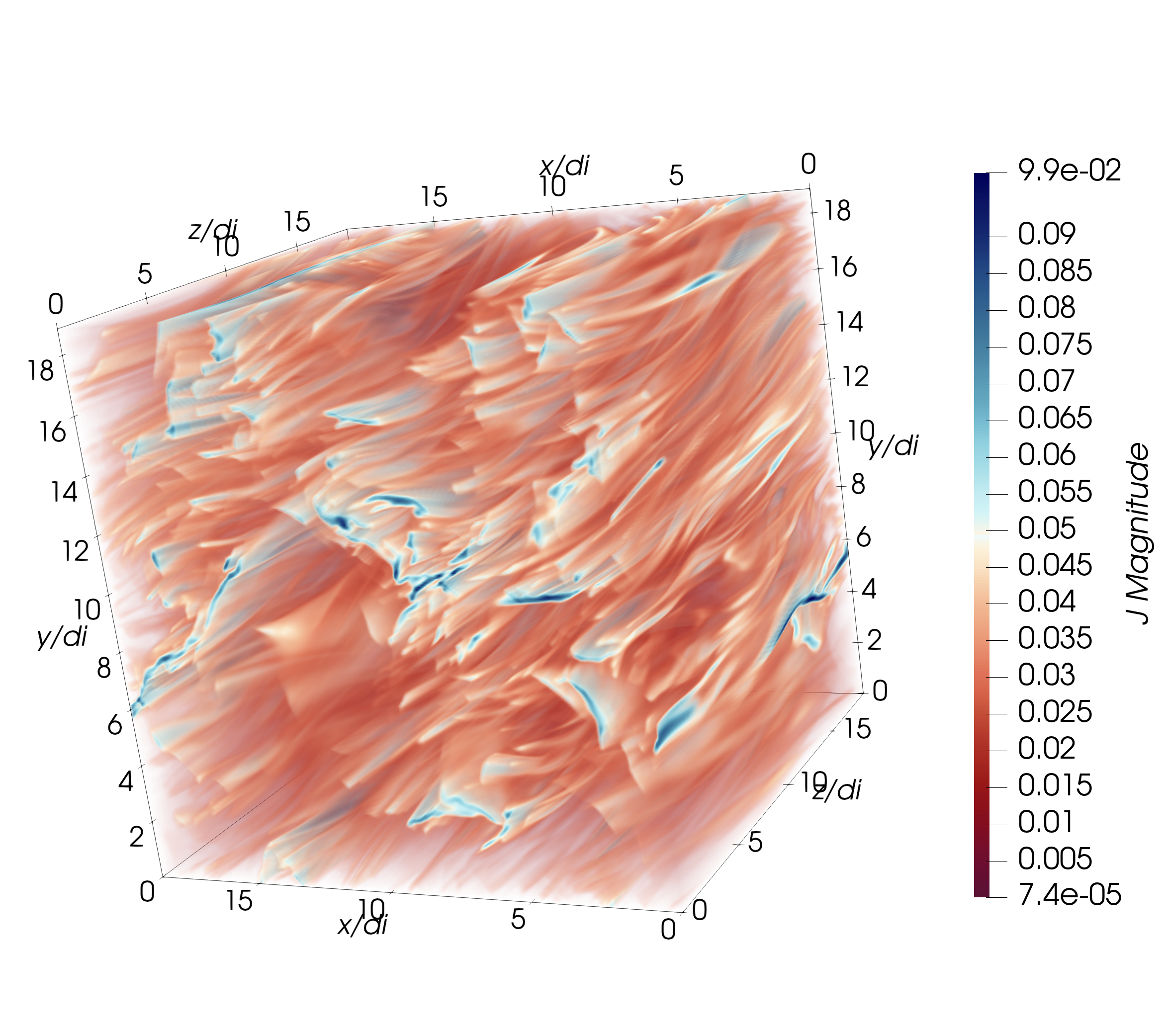} 
  \includegraphics[width=0.325\textwidth]{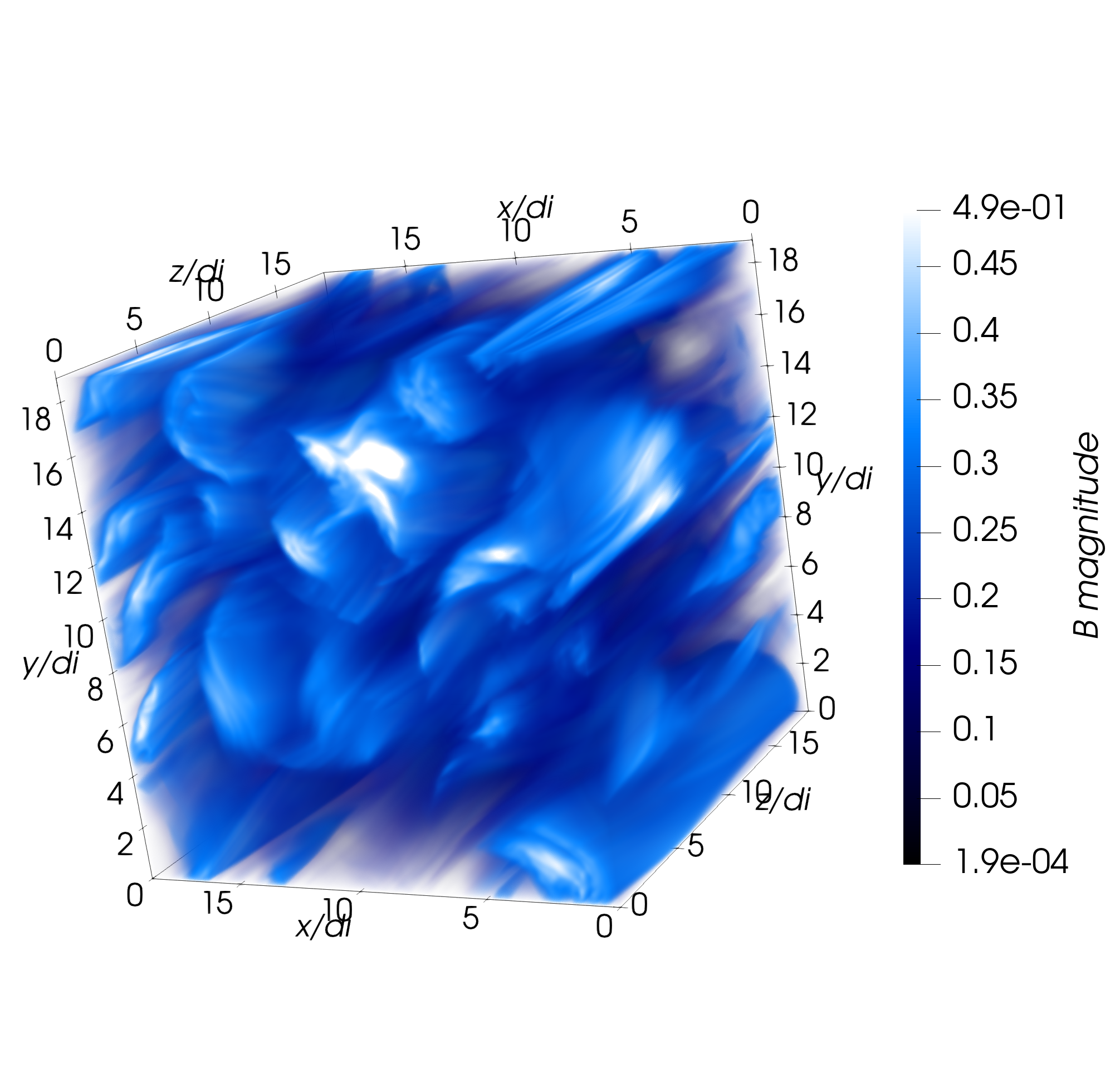} 
  \includegraphics[width=0.325\textwidth]{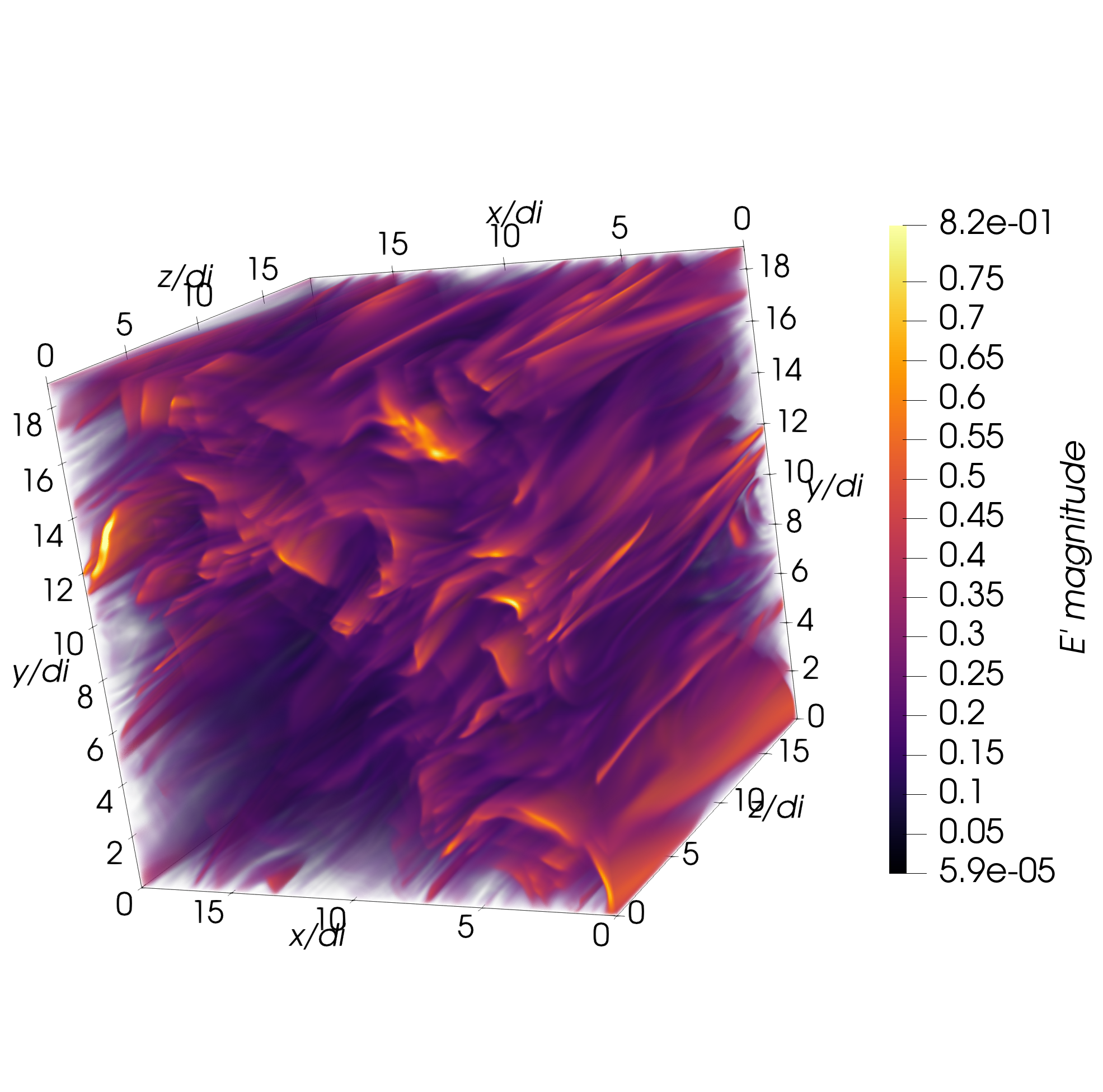} 
  \includegraphics[width=0.325\textwidth]{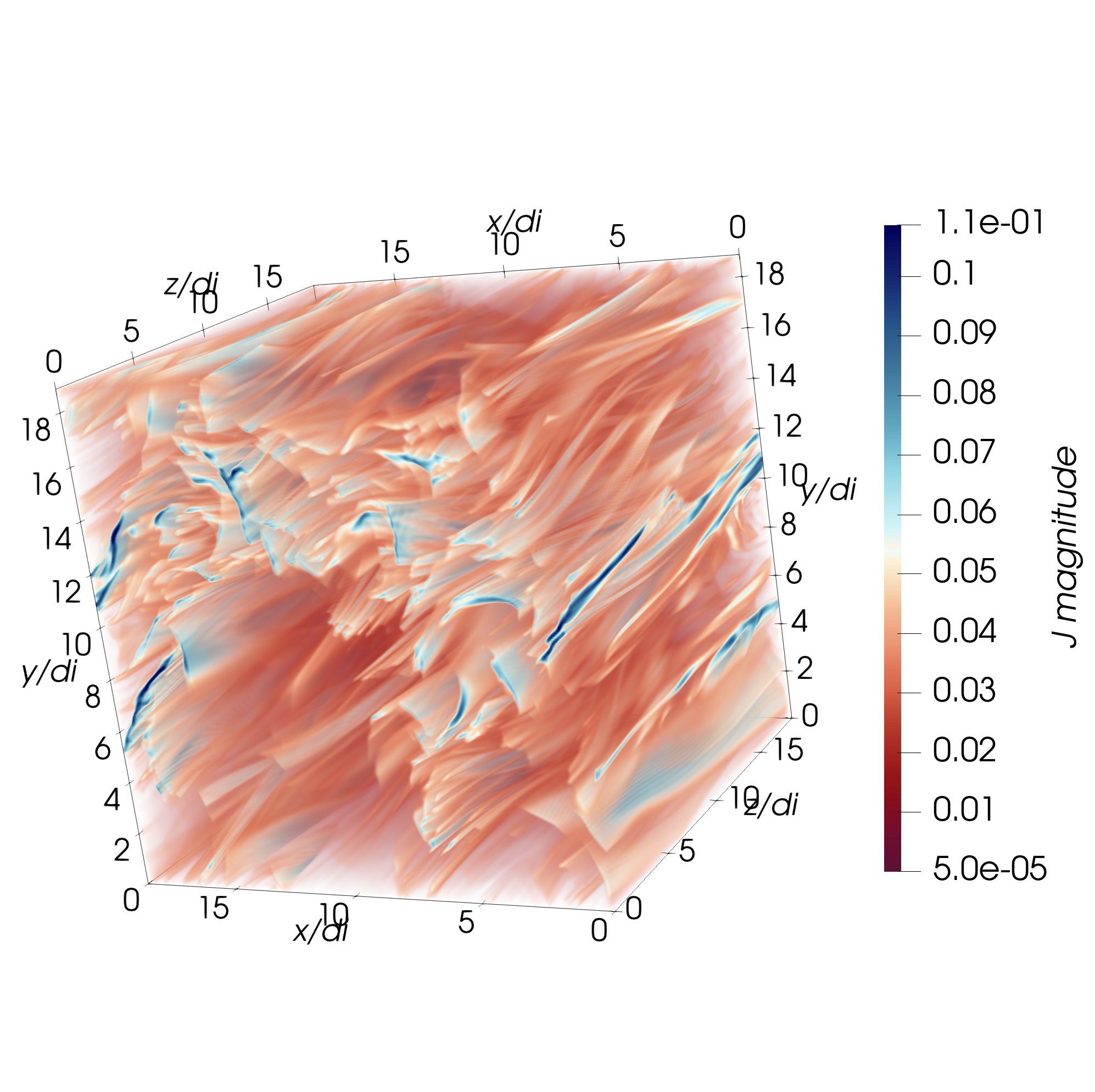} 
  \caption{3D rendering of the magnitude of $\delta \boldsymbol{B}$ (left column), $\boldsymbol{E}'=\boldsymbol{E}+\boldsymbol{u}_e\times\boldsymbol{B}$ (center column), and $\boldsymbol{J}=\nabla\times\boldsymbol{B}$ (right column), at the turbulent peak activity for $\beta_{\rm i}=0.25$ (top row), $\beta_{\rm i}=1$ (middle row), and $\beta_{\rm i}=4$ (bottom row).}
  \label{fig:3D_bperp}
\end{figure*}

\subsection{Caveats, limitations, and future improvements}\label{subsec:limitations}

From the model point of view, the hybrid-kinetic approximation adopted for the current simulations clearly neglects several electrons' kinetic physics, such as electron finite-Larmor-radius (eFLR) effects and electron Landau damping (eLD), while still retaining electron-inertia effects. This limits this study to regimes with low electron beta, $\beta_{\rm e}\ll1$, so that the electron gyro-radius $\rho_{\rm e}$ is sufficiently smaller than the electron inertial length $d_{\rm e}$ (e.g., $\beta_{\rm e}=0.1$ in our simulations; see Section~\ref{subsec:sim_setup}). While this regime seems to be suitable for electron-only reconnection in the Earth's magnetosheath~\citep[e.g.,][]{ChenBoldyrevAPJ2017,StawarzAPJL2019}, it could be of interest to explore also the $\beta_{\rm e}\gtrsim1$ regime in the context of other astrophysical environments. 
Nonetheless, in the context of the observed regimes for electron-only reconnection, the hybrid-kinetic model still offers a suitable compromise between a computationally much more expensive full-kinetic approach and keeping the whole ion-kinetic physics (and associated heating processes) that would not be included in other, less computationally demanding, reduced-kinetic models such as, for instance, gyro-kinetics.
Furthermore, the hybrid-kinetic approach is partially justified by the idea that across the ion scales and in the sub-ion range of scales above the electron Larmor radius, electron-heating processes (e.g., eLD) are negligible with respect to other ion-heating mechanisms. While this may not be entirely true when comparing eLD with ion Landau damping (iLD) in a moderately collisional gyrokinetic plasma~\citep[e.g.,][]{ToldPRL2015}, it seems to be well justified when stochastic heating and cyclotron damping are taken into account, since they all largely dominate over iLD in the ions' turbulent heating~\citep[e.g., see][and references therein]{ArzamasskiyAPJ2019,CerriAPJ2021,Arz23}.
Finally, a way to partially overcome this caveat is to include, still within a hybrid-kinetic description, more refined electron models with, for instance, anisotropic electron pressure and Landau-fluid closures accounting for eLD~\citep[e.g.,][]{FinelliAA2021}.

From the numeric point of view, our simulations are limited by computational capabilities and available resources. This affects the maximum size of the simulation domain, its small-scale resolution, and the need to adopt a reduced mass ratio (i.e., in our simulations, depending on the ion beta parameter $\beta_{\rm i}$, this is a six-dimensional phase-space domain of maximum sizes $L_{j}=6\pi d_{\rm i}$ and $|v_{j}|\leq 7\, v_{\rm th,i}$ (with $j=x,y,z$) discretized with up to $256^3\times 57^3$ grid points, respectively, and adopting a mass ratio $m_{\rm i}/m_{\rm e}=100$; see Section~\ref{subsec:sim_setup}). Considering a realistic mass ratio could affect the reconnection rates in the simulations. Specifically, in the EMHD regime, the linear growth rate for marginally unstable tearing modes is proportional to $(m_{\rm e}/m_{\rm i})$  (\cite{BulanovPOF1992}), resulting in a slower reconnection rate if more realistic mass ratio values were considered. On the other hand, the larger scale separation would allow for better decoupling of the species dynamics, potentially highlighting clearer electron-only reconnection events.
Ideally, one would like to slightly increase both the simulation box and, possibly, consider also the case of an anisotropic domain $L_\|>L_\perp$ to further explore the role of different $k_{\rm inj}d_{\rm{i}}$, $k_{\rm inj}\rho_{\rm{i}}$, and $k_{\rm\|,inj}/k_{\rm\perp,inj}$; the role of a mean field by varying $\delta B_{\rm inj}/B_0$ is also something that needs to be explored in the future. Achieving better resolutions will also allow to further extend the sub-ion range with more realistic mass ratios and to better describe the sub-electron-scale dynamics involved in electron-only reconnection. However, these improvements cannot be reached at the present time and will have to await significantly better computational resources. Nevertheless, the present work still represents the first attempt to extend previous studies of electron-only reconnection in hybrid-kinetic plasma turbulence from 2.5D to 3D, and with different ion-beta regimes (namely, $\beta_{\rm i}=0.25$, $1$, $4$).

\section{Results}\label{sec:results}

We let the initial condition described in Section~\ref{subsec:sim_setup} freely decay into a turbulent state. Fully developed turbulence, denoted by a peak in the rms current density $J_\|^{\rm rms}$ (not shown), is achieved at $t \approx 14\, \Omega_{\rm c,i}^{-1}$ for the case $\beta_{\rm i} = 0.25$,  at $t\approx 11\,\Omega_{\rm c,i}^{-1}$ for the case $\beta_{\rm i} = 1$ and at $t \approx 13\,\Omega_{\rm c,i}^{-1}$ for the case $\beta_{\rm i} = 4$.
Throughout the fully developed turbulent state, the rms of magnetic-field fluctuations remain roughly stable at a level $\delta B_{\rm rms}/B_0\approx 0.3$ (not shown).
In Figure~\ref{fig:3D_bperp} we show a three-dimensional rendering of the magnetic-field-fluctuation modulus $|\delta\boldsymbol{B}|$ (left column), of the modulus of the electric field in the electron frame $|\boldsymbol{E}'|=|\boldsymbol{E}+\boldsymbol{u}_e\times\boldsymbol{B}|$ (center column), and of the current-density modulus $|\boldsymbol{J}|$ (right column) at the time of the peak for the three simulations, $\beta_{\rm i}=0.25$ (top row), $\beta_{\rm i}=1$ (middle row), and $\beta_{\rm i}=4$ (bottom row).

\begin{figure}[t]
  \centering
  \includegraphics[width=0.325\textwidth]{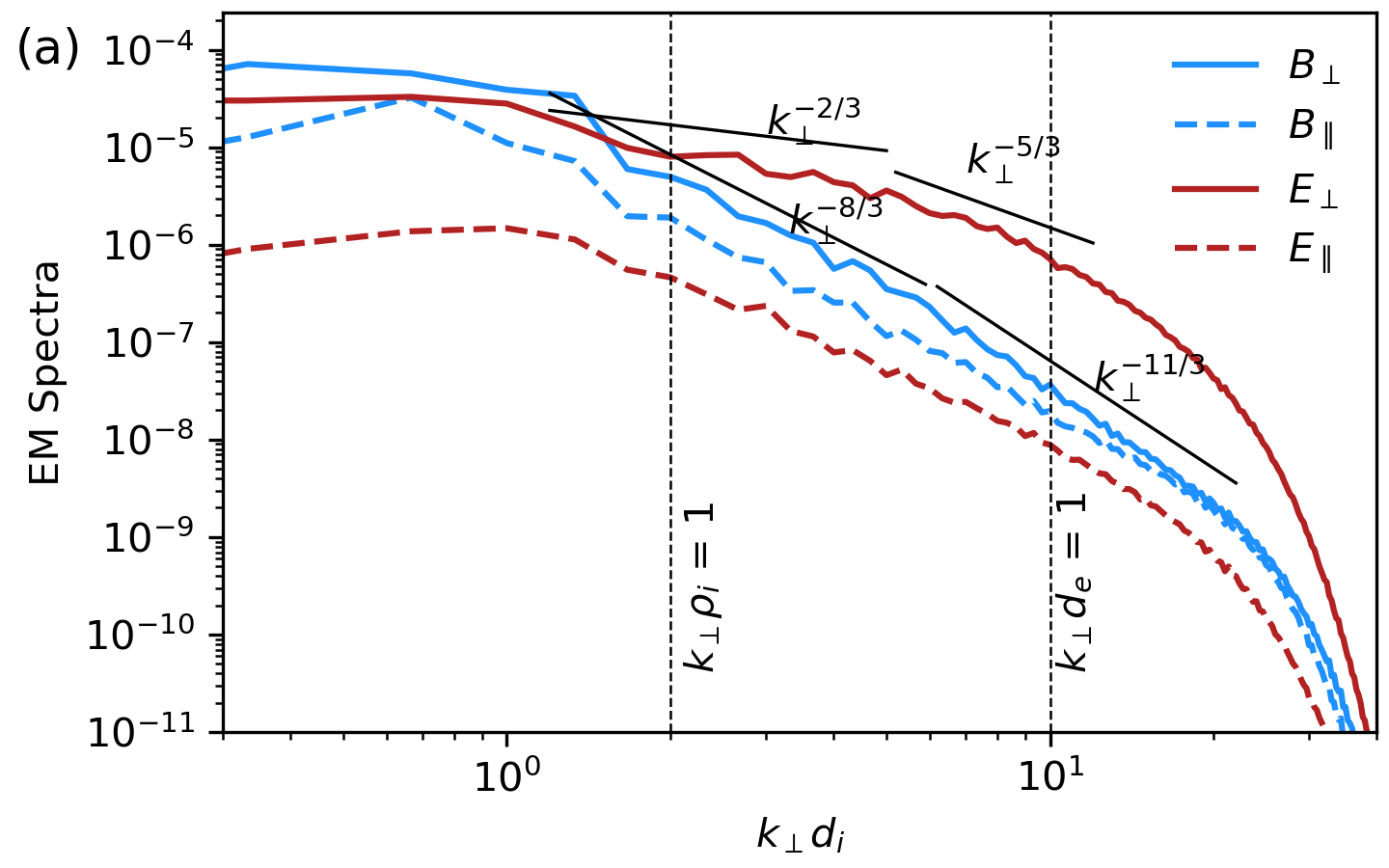} 
  \includegraphics[width=0.325\textwidth]{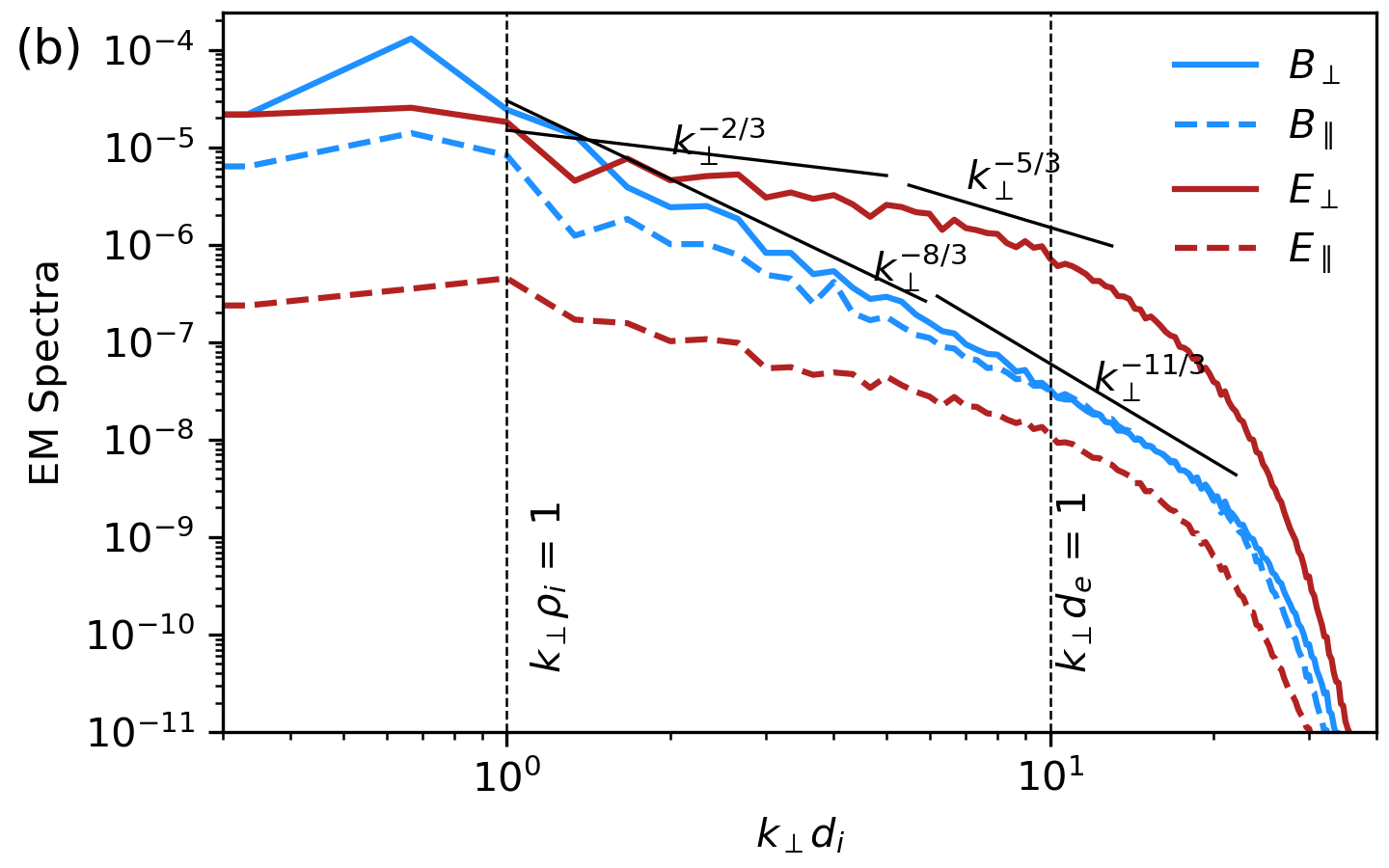}
    \includegraphics[width=0.325\textwidth]{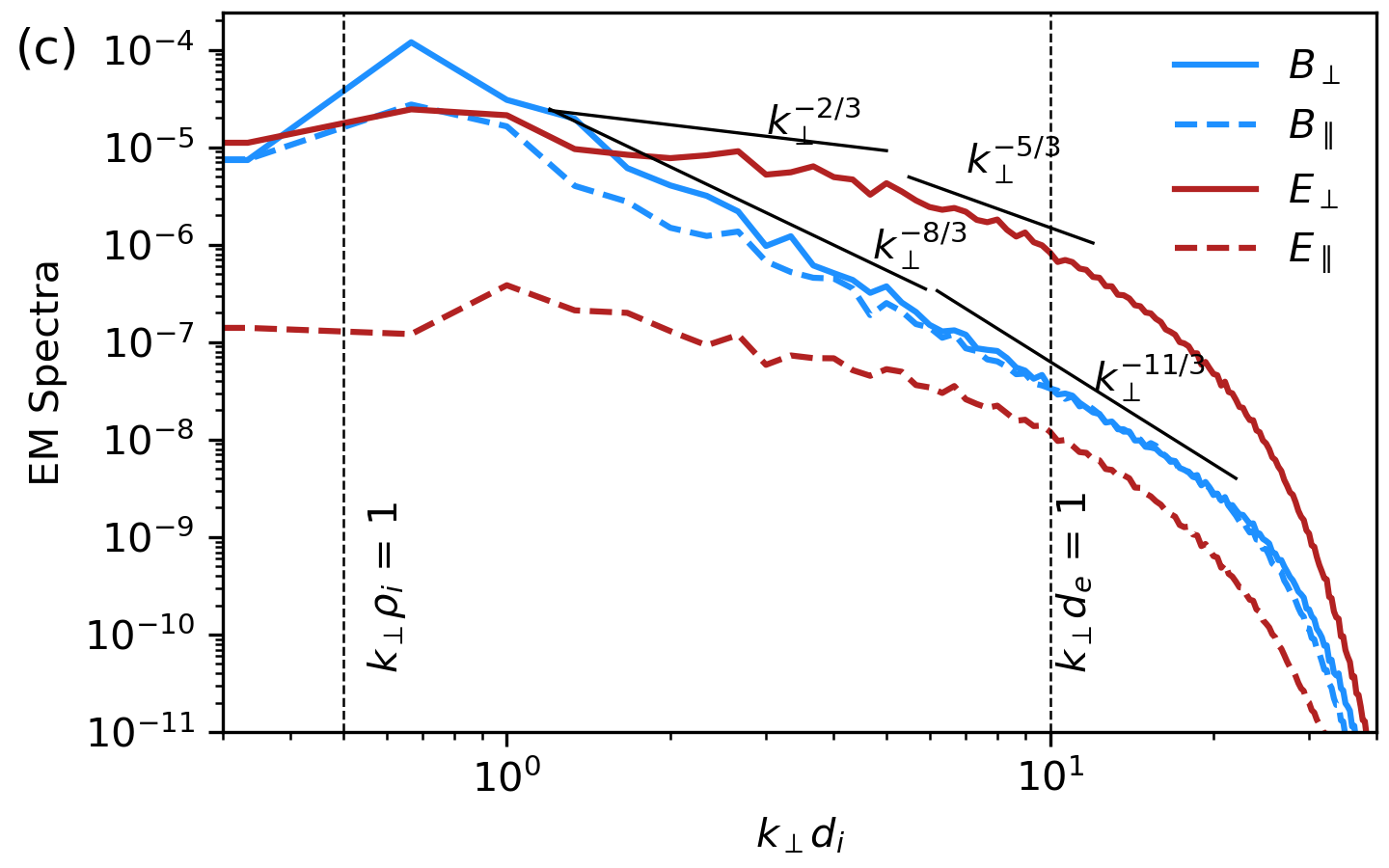} 
  \caption{Spectra of electromagnetic fluctuations versus $k_\perp$ for $\beta_{\rm i} = 0.25$ (panel (a)), $\beta_{\rm i} = 1$ (panel (b)) and $\beta_{\rm i} = 4$ (panel (c)) averaged around the turbulent peak activity. Relevant slopes are also given as a reference.}
  \label{fig:Spectra_EB}
\end{figure}

\subsection{Spectral properties}\label{subsec:spectra}

It is instructive to first focus our attention on the spectral properties of fluctuations. This will help to clarify which type of turbulence develops in our simulations, and if there are differences related to a specific $\beta$ regime.

In Figure~\ref{fig:Spectra_EB}, we show the reduced (i.e., $k_\|$-averaged) power spectrum of electromagnetic fluctuations, $\delta B_\perp$, $\delta B_\|$, $\delta E_\perp$, and $\delta E_\|$, versus the perpendicular wavevector $k_\perp$, time-averaged around the peak activity of turbulence. We remind the reader that here the perpendicular direction is defined with respect to $\boldsymbol{B}_0$ rather than to a local-in-scale mean field $\langle\boldsymbol{B}\rangle_k$, i.e., $k_\perp=(k_x^2+k_y^2)^{1/2}$. While this should not affect significantly the fluctuations' properties in the perpendicular direction for our level of fluctuations in the quasi-steady state, we notice that it would affect more significantly their properties in the parallel direction (i.e., the properties inferred using $k_z$ instead of a local $k_\|$). For this reason, the following analysis will focus only on $k_\perp$ spectra. Furthermore, due to the limited box size of the simulations and of the initial condition, we cannot really draw conclusions about the spectrum above ion scales, so our focus will be on sub-ion scales ($k_\perp d_{\rm i}>1$). 

\begin{figure}[t]
  \centering
  \includegraphics[width=0.45\textwidth]{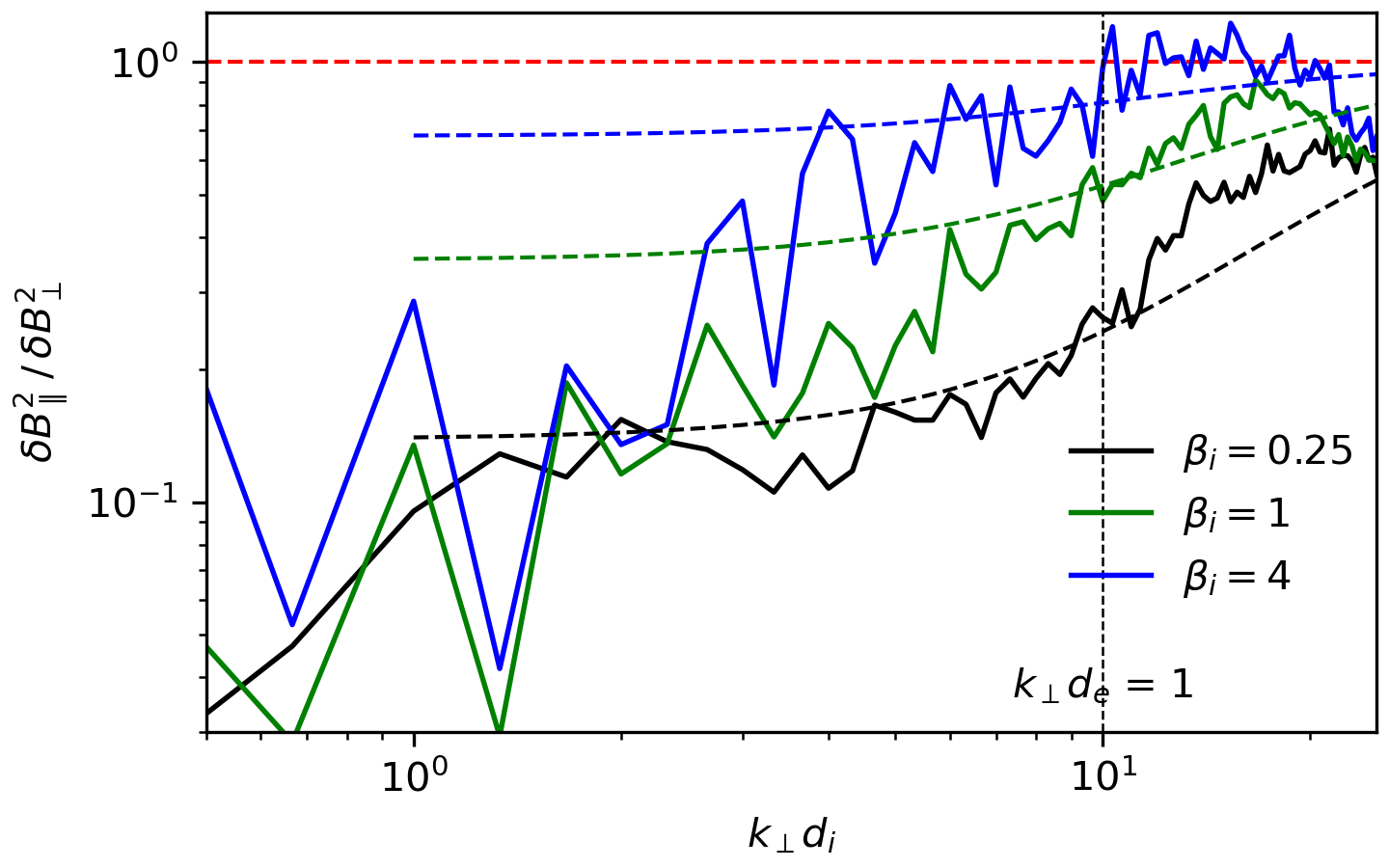}
  \includegraphics[width=0.45\textwidth]{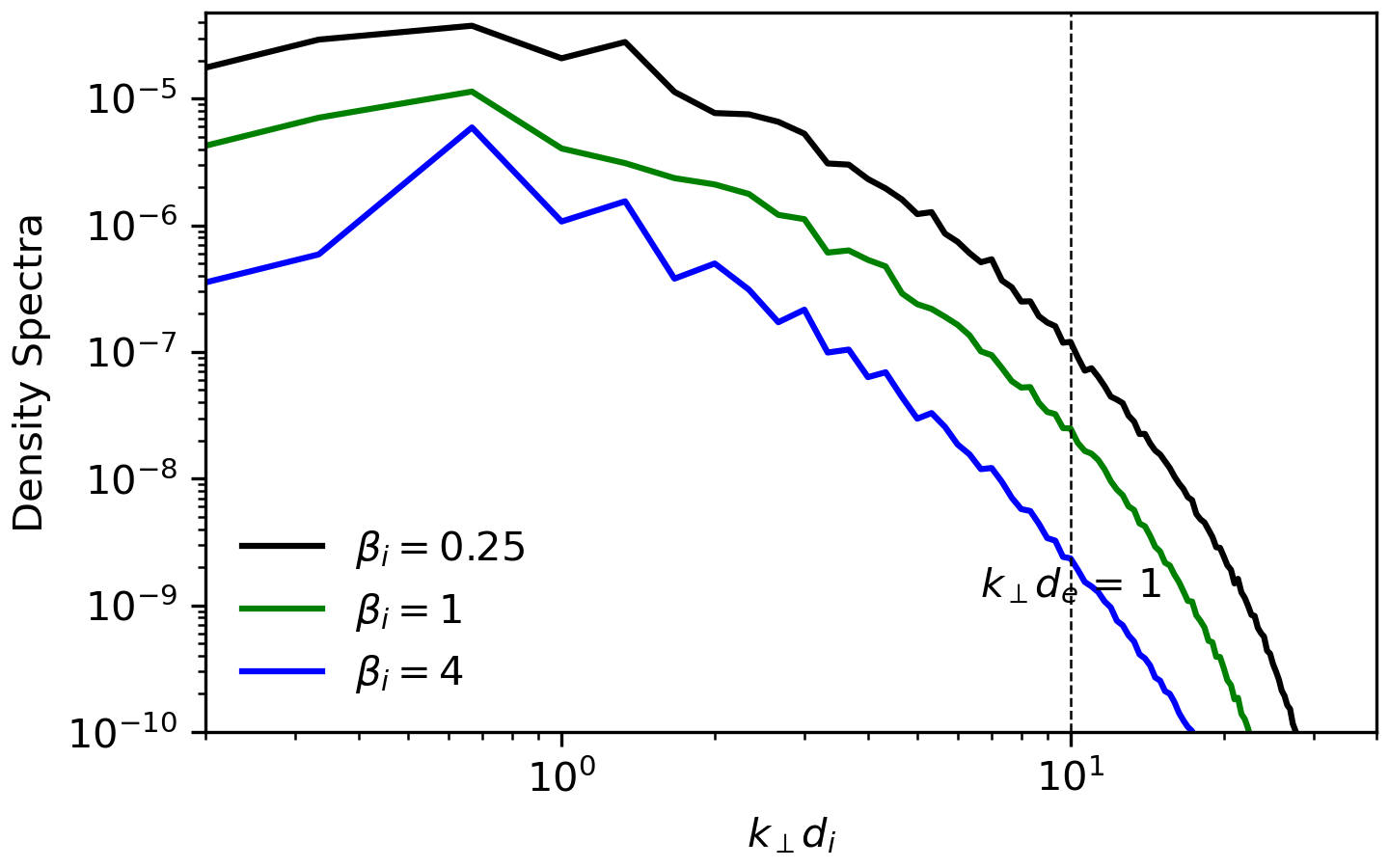}
  \caption{Left: scale-dependent magnetic compressibility, $\delta B_{\|}^2/\delta B_{\perp}^2$, for the three simulations (see legend). The dashed curves of the corresponding colors represent the analytical prediction given in \eqref{compikaw} for different $\beta_{\rm i}$, while the horizontal red dashed line corresponds to the approximated (scale- and $\beta_{\rm i}$-independent) prediction for whistler fluctuations given in \eqref{compiww}. Right: spectrum of density fluctuations versus $k_\perp d_{\rm i}$. }
  \label{fig:MagCompr}
\end{figure}

In all $\beta_{\rm i}$ regimes, the sub-ion-scale spectrum of $B_\perp$ fluctuations consistently exhibits a $\propto k_\perp^{-8/3}$ range above the electron scales (i.e., in the wavevector interval  $1\lesssim k_\perp d_{\rm i}\lesssim 6.5$). This slope, which is in agreement with previous 3D kinetic simulations of sub-ion-scale turbulence~\citep[e.g.,][]{ToldPRL2015,CerriAPJL2017,FranciAPJ2018,GroseljPRL2018,RoytershteynAPJ2019} and with solar-wind/magnetosheath observations~\cite[e.g.,][]{AlexandrovaPRL2009,AlexandrovaSSRv2013,SahraouiPRL2009,SahraouiRMPP2020,KiyaniRSPTA2015,ChenBoldyrevAPJ2017,StawarzAPJL2019}, can be explained in terms of intermittency corrections to a cascade of KAW-like fluctuations~\citep{BoldyrevPerezAPJL2012}. 
Across and below electron scales,  $0.7\lesssim k_\perp d_{\rm e}\lesssim 2$ (corresponding to the range  $6.5\lesssim k_\perp d_{\rm i}\lesssim 20$), the $\delta B_\perp$ spectrum steepens and becomes qualitatively consistent with a $\propto k_\perp^{-11/3}$ power law.  It is interesting to note that this steepening occurs slightly above the $k_\perp d_{\rm e}\approx1$ scale.
The spectrum of $\delta B_\|$ follows qualitatively that of $\delta B_\perp$, being slightly shallower at  $k_\perp d_{\rm i}< 7$ (roughly consistent with a $-7/3$ slope). This slight discrepancy in the spectral slope is related to the fact that $\delta B_\|^2<\delta B_\perp^2$ at larger scales, while the power in the two magnetic-field components will eventually reach equipartition at sub-electron scales (see the analysis of magnetic compressibility in Figure~\ref{fig:MagCompr} and associated discussion below).
The electric-field spectrum at sub-ion scales initially exhibits a shallow power law, $\propto k_\perp^{-2/3}$, that steepens as it approaches $k_\perp d_{\rm e}\sim1$ (roughly $\propto k_\perp^{-5/3}$), before eventually being exponentially dissipated by numerical filters. The $-5/3$ slope appearing roughly in the same range
where the magnetic-field spectrum steepens to $-11/3$ is consistent with a transition to either inertial kinetic-Alfv\'en-wave (IKAW) or inertial whistler-wave (IWW) turbulence~\citep{ChenBoldyrevAPJ2017,RoytershteynAPJ2019}.
We remind the reader that the scaling for IKAW and IWW fluctuations are valid in the limit of negligible eFLR corrections~\citep[see, e.g., also][]{PassotJPP2017,PassotPOP2018}, which is consistent with the hybrid-kinetic approximation that we are employing. Moreover, the choice $\beta_{\rm e}=0.1$ ensures that eFLR effects would be anyway confined to the dissipation range of our simulations (see Section~\ref{subsec:sim_setup}). Therefore, within the range of scales that is being investigated here, we would not expect significant deviations from our results even if a full-kinetic model was to be employed~\citep[see, e.g.,][]{RoytershteynAPJ2019}. It would be interesting to investigate finite-$k_\perp\rho_{\rm e}$ effects on this type of cascade, but this is out of the scope of this work and would require a model beyond the basic hybrid-kinetic one that is adopted here (see discussion in Section~\ref{subsec:limitations}).
Since the electromagnetic spectrum alone cannot distinguish between these two possible regimes, and in general there is virtually no difference between the spectra in Figure~\ref{fig:Spectra_EB} at sub-ion scales for different $\beta_{\rm i}$, it is necessary to look also at other quantities.

In order to further assess the small-scale behaviour of turbulent fluctuations, in Figure~\ref{fig:MagCompr} we report the scale-dependent magnetic compressibility, $\delta B_\|^2/\delta B_\perp^2$ (left panel), and the spectrum of density fluctuations (right panel) for the three simulations.

A transition between KAW and IKAW regimes at sub-ion scales (and for $\tau=T_{\rm i}/T_{\rm e}\gg1$) would reflect in the scale-dependent magnetic compressibility following the relation~\citep{ChenBoldyrevAPJ2017}
\begin{equation}
\frac{\delta B_\parallel^2}{\delta B_{\perp}^2} = \frac{1 + k_{\perp}^2 d_{\rm e}^2}{1 + \frac{2}{\beta_{\rm i}} + k_{\perp}^2 d_{\rm e}^2}\,,\label{compikaw}
\end{equation}
which shows that such ratio increases from its scale-independent value $\delta B_\|^2/\delta B_\perp^2=1/(1+2/\beta_{\rm i})$ typical of standard KAW fluctuations due to finite-$k_\perp d_{\rm e}$ effects in the IKAW regime (see $\beta_{\rm i}$-dependent dashed lines in the left panel of Figure~\ref{fig:MagCompr}).  In calculating the theoretical curves with equation (4), we used the value of $\beta_i$ measured at the time chosen to plot the magnetic compressibility.  This transition can be attributed to a shift in the dominant drift mechanism in the perpendicular plan. 
On the other hand, the magnetic compressibilty in the IWW regime is $\beta_{\rm i}$-independent and given by~\citep{ChenBoldyrevAPJ2017}
\begin{equation}
\frac{\delta B_\parallel^2}{\delta B_{\perp}^2} = \frac{1 + k_\|^2 d_{\rm i}^2 + k_{\perp}^2 d_{\rm e}^2}{k_\|^2 d_{\rm i}^2}\approx 1\,,\label{compiww}
\end{equation}
where the last approximation is valid for sub-ion-scale fluctuations with $k_\|^2d_{\rm i}^2\gg k_\perp^2d_{\rm e}^2\sim 1$ (i.e., for fluctuations that are not extremely oblique, $k_\|/k_\perp\gg (m_{\rm e}/m_{\rm i})^{1/2}$, and with perpendicular wavelength in the electron-scale range, $k_\perp d_{\rm e}\sim1$). This scale-independent approximation for the whistler magnetic compressibility is indeed appropriate for the wavenumber range of our simulations. Moreover, since a cascade of IKAW fluctuations becomes progressively less anisotropic~\citep[following a relation $k_\|\propto k_\perp^{5/3}$;][]{ChenBoldyrevAPJ2017}, it is possible that when a cascade of very oblique KAW fluctuations turns to the IKAW regime, its anisotropy may get reduced enough to further transition to IWW turbulence. From the behaviour of the magnetic compressibility in the left panel of Figure~\ref{fig:MagCompr}, this seems to be the case for our simulations.
 
For the case $\beta_{\rm i} = 0.25$, the magnetic compressibility aligns with  KAW/IKAW relation \eqref{compikaw} in the region  $1 \lesssim k_\perp d_{\rm i} \lesssim 8$ and increases towards unity more rapidly than what expected for IKAW fluctuations for $k_\perp d_{\rm e}\gtrsim1$, possibly denoting a (partial) transition to IWW turbulence at electron scales. 
A similar qualitative behaviour is observed also for the other two simulations, $\beta_{\rm i}=1$ and $\beta_{\rm i}=4$, although the reduced difference between the scale-dependent IKAW curves for these $\beta_{\rm i}$ values and the scale-independent IWW prediction \eqref{compiww} makes this distinction less clear.
However, this behaviour becomes more clear if we consider the spectrum of density fluctuations (right panel in Figure~\ref{fig:MagCompr}). In fact, in the KAW/IKAW regime, density fluctuations are related to the parallel magnetic-field fluctuations by the relation $(\delta B_\|/B_0)^2 = (\beta_{\rm i}^2/4)(\delta n/n_0)^2$ and thus they cannot be neglected ($\delta n/n_0\lesssim1$ is finite). On the other hand, in whistler turbulence (and, in general, in the EMHD regime, for which ions' response is negligible) the density fluctuations are negligibly small, $\delta n/n_0\ll1$. In our simulations the density spectrum indeed drops dramatically at sub-ion scales, especially at $\beta_{\rm i}=4$ (blue curve in the right panel of Figure~\ref{fig:MagCompr}). This behaviour supports the idea that the ions' response is more negligible (and thus the EMHD regime is more easily achieved at sub-ion scales) when the energy is injected close to the ion Larmor radius $\rho{\rm i}$ rather than to the ion inertial length $d_{\rm i}$. As a result, we expect that the electron-only reconnection regime is more easily achieved in our $\beta_{\rm i}=4$ simulation, and, in general, in high-$\beta_{\rm i}$ turbulence.

\begin{figure}[t]
  \centering
\includegraphics[scale=0.485]{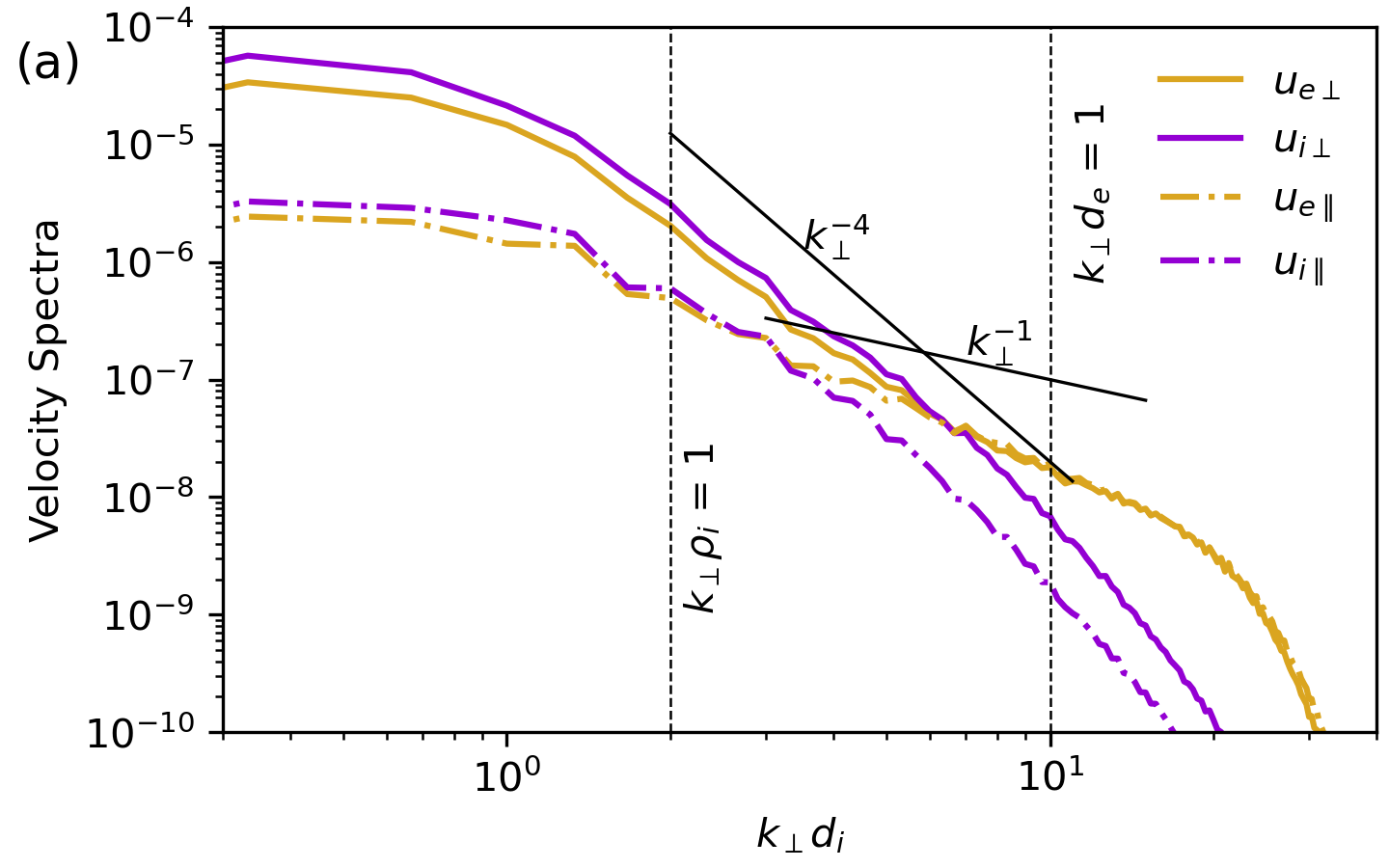}
  \includegraphics[scale=0.485]{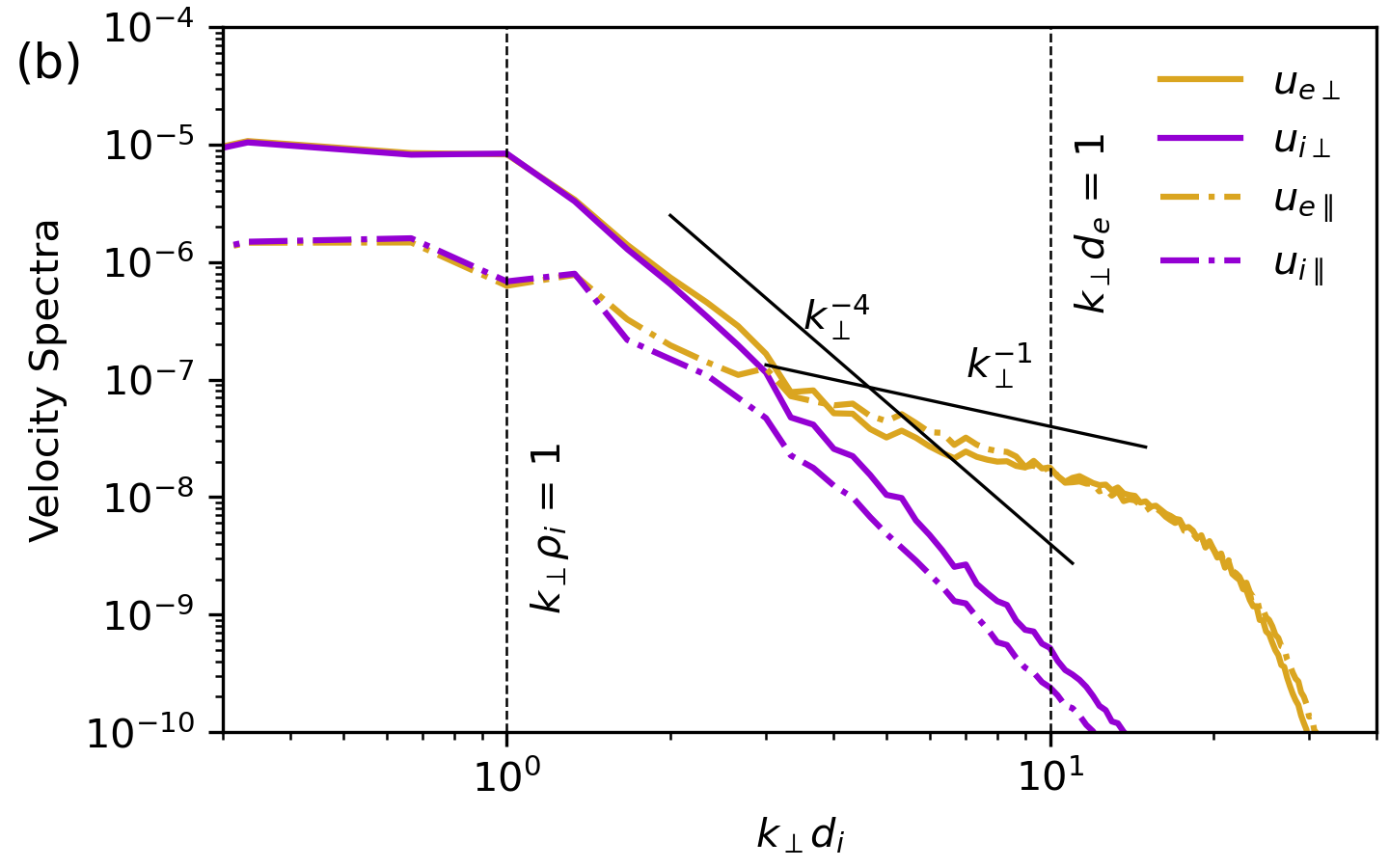}
  \includegraphics[scale=0.485]{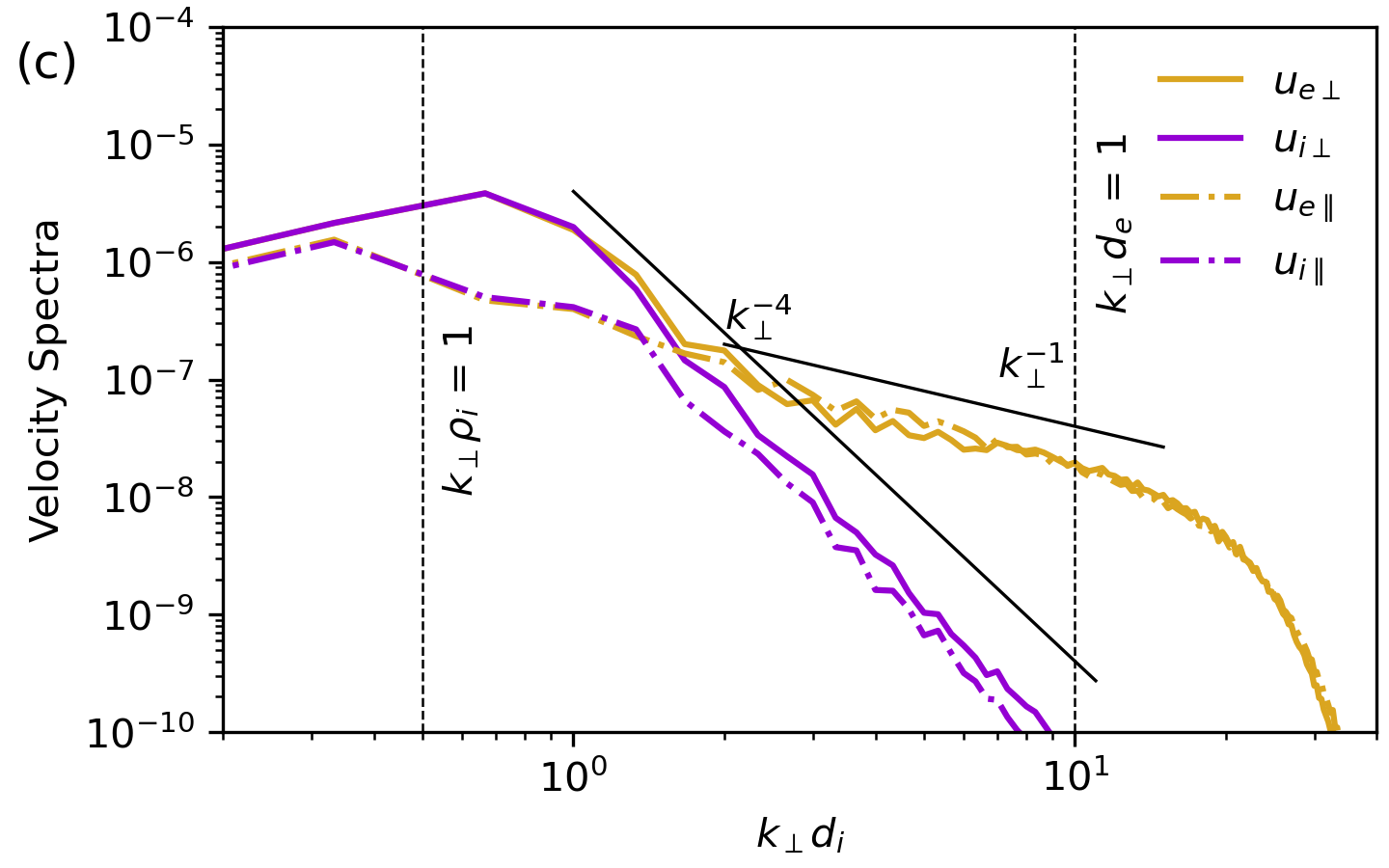} 
  \caption{Spectra of perpendicular solid lines and parallel (dot-dashed lines) electron (gold) and ion (purple) velocities for $\beta_{\rm i} = 0.25$ (panel (a)), $\beta_{\rm i} = 1$ (panel (b)) and $\beta_{\rm i} = 4$ (panel (c)).}
  \label{fig:Spectra_ueperp_uiperp}
\end{figure}

The scale-dependent species' response at small scales is addressed in Figure \ref{fig:Spectra_ueperp_uiperp}, where we report the spectra of the parallel and perpendicular ion and electron flows. Consistently with results from previous 2.5D hybrid-kinetic simulations~\citep{CerriJPP2017}, also our 3D simulations clearly show that the ion flow dramatically drops at $k_\perp\rho_{\rm i}\gtrsim 3$, after which electron flows dominate the sub-ion-scale dynamics. This means that as the ion beta increases from $\beta_{\rm i}=1/4$ to $\beta_{\rm i}=4$ (and thus does the ratio between the ion Larmor radius and the ion inertial length, from $\rho_{\rm i}/d_{\rm i}=1/2$ to $\rho_{\rm i}/d_{\rm i}=2$), a larger portion of the sub-ion range in our simulations lacks of ion response.
As far as the reconnection regime is concerned, for the $\beta_{\rm i}=4$ case this will likely reflect into more prominent electron-only reconnection events where electron outflows are not being accompanied by ion outflows (i.e., a reconnection region with an EDR and no IDR). 
It is crucial to point out that it is the presence of a  non-negligible guide field (for example, $\delta B_{\rm rms}/B_0\simeq0.3$ as in our case), that makes this transition depending on the ion gyroscale $\rho_{\rm i}$ (rather than on $d_{\rm i}$).

\subsection{Identification of 3D reconnection events}\label{subsec:3Drec_events}

Detecting reconnection events in 3D poses a challenging task. 
In this study, we make use of four specific signatures to effectively identify these events within our simulations. 

The first and arguably the most reliable indicator for detecting reconnection in 3D space is the presence of an important magnetic-field-aligned electric field within an active reconnection regions. To identify EDR, an established criterion is to examine areas where the component of the electric field parallel to the reconnecting plane, in the electron-fluid frame, is nonzero: $E_{\parallel_{\text{RP}}}'= E_{\parallel_{\text{RP}}} + (\boldsymbol{u}_{\rm e} \times \boldsymbol{B})_{\parallel_{\text{RP}}} \neq 0$ \citep{Dau06, Fuj06, Mun23}. (Here, we denoted quantities parallel to the reconnection plane with the symbol $"{\parallel_{\text{RP}}}"$. We recall that, in the rest of the paper, the symbol ``$\|$'' means parallel to the mean-field direction, $\boldsymbol{B}_0=B_0\hat{\boldsymbol{z}}$.) In this work, although the current appears to predominantly form small structures in the $xy$ plane and the magnetic flux ropes seem more elongated along the $z$ direction (see the 3D rendering in Fig \ref{fig:3D_bperp}), we do not rule out the possibility of field lines reconnecting in a plane inclined relative to the $xy$ plane. Consequently, we look for regions where the total magnitude of $\boldsymbol{E}'=\boldsymbol{E}+\boldsymbol{u}_{\rm e} \times \boldsymbol{B}$  becomes significant, (in particular, for our simulations we find that a threshold $|\boldsymbol{E}'|\geq 0.6$ provides a reliable criterion for this purpose). A second indicator of magnetic reconnection is the presence of field-aligned current sheets, which leads us to check the parallel current density  $J_\|$. A third essential signature of reconnection is the reversal of the magnetic field components in the reconnection plane. A fourth distinctive feature of a reconnection event is the presence of inflows and outflows directed toward and outward the reconnection point, respectively. Ideally, to identify electron-only reconnection sites, we aim to  observe cases characterized by the absence of ion outflows, meaning a reconnection event occurring without an IDR. Still, electron outflows from an EDR will be present in any type of reconnection site.

To provide an estimate of the current sheet dimensions across the three different simulations, we measured their widths and lengths~\citep[see, e.g.,][and refereces therein for a discussion of different methods]{SistiAA2021}. For the width, we fitted a Gaussian profile along the x or y direction that passes through the maximum current value in a z-plane where the current reaches its peak. The width is then defined as the Full Width at Half Maximum (FWHM) of the Gaussian. Once the width is determined, the length of the current sheet is defined as twice the distance between the position where the maximum current value is attained and the position where the current reaches the FWHM value of the Gaussian, corresponding to the edge of the current sheet along its length. 
The simulations for $\beta_{\rm i} = 0.25$ and $\beta_{\rm i} = 1$ (cases for which  $\rho_{\rm i}\leq d_{\rm i}$) appear to yield similar results concerning the the geometry of current sheets. For these simulations, current sheets have lengths of approximately $L_{\rm CS} \sim 3.2 d_{\rm i}$ with a width of $\delta_{\rm CS} \sim 0.4 d_{\rm i} = 4 d_{\rm e}$. On the other hand, in the $\beta_{\rm i}=4$ regime (for which $\rho_{\rm i}> d_{\rm i}$), the current sheets appear to always be thinner and shorter with respect to the two lower-$\beta_{\rm i}$ cases, namely $L_{\rm CS} \sim 1.81 d_{\rm i}\sim 3.6\rho_{\rm i}$ and $\delta_{\rm CS} \sim 0.2 d_{\rm i} \sim 0.4\rho_{\rm i} = 2 d_{\rm e}$. 
This seems to be in line with a picture where the overall size of the kinetic-scale current sheets are determined by the largest of the ion scale $\lambda_{\rm i,max}\sim \max(d_{\rm i},\rho_{\rm i})$, i.e., $L_{\rm CS}\sim 3$--$4\,\lambda_{\rm i,max}$ and $\delta_{\rm CS}\sim 0.4\,\lambda_{\rm i,max}$, and their aspect ratio that is almost always consistent with a sort of limiting value $L_{\rm CS}/\delta_{\rm CS} < 10$.
This aligns with a recent analysis of electron-only tearing mode conducted by \cite{Mal20}, which shows a transition from the ion-coupled regime to a the electron-only reconnection regime when the aspect ratio of current sheet is below $10$ and the thickness of the current sheet is below the ion-sound Larmor radius, $\rho_s$.

Then, what matters for the occurrence of electron-only reconnection is how much the ions can couple with the dynamics of these current sheets, which indeed is determined by how their Larmor radius $\rho_{\rm i}$ compare with the current-sheet sizes. As a result, we expect to find that ions decouple from the electron fluid and kinetic-scale current sheets more easily at high $\beta_{\rm i}$. 
Clearly, the properties of kinetic-scale current sheets highlighted above may be affected by the reduced mass ratio $m_{\rm i}/m_{\rm e}=100$ employed in our simulations, as well as to the limited box size and span in plasma beta (see discussion in Section~\ref{subsec:limitations}). A systematic and more detailed study of these current-sheets' properties requires larger 3D kinetic simulations able to explore more realistic mass ratios and a wider spread in $\beta_{\rm i}$, which is beyond the scope of this work and will have to await better computational capabilities. Nevertheless, the properties of kinetic-scale current sheets observed in our simulations seem to be qualitatively consistent with previous kinetic simulations and in-situ observations~\citep[e.g.,][]{StawarzAPJL2019,Sta22,VegaMNRAS2023}.

\begin{figure}[t]
  \centering
  \includegraphics[scale=0.13]{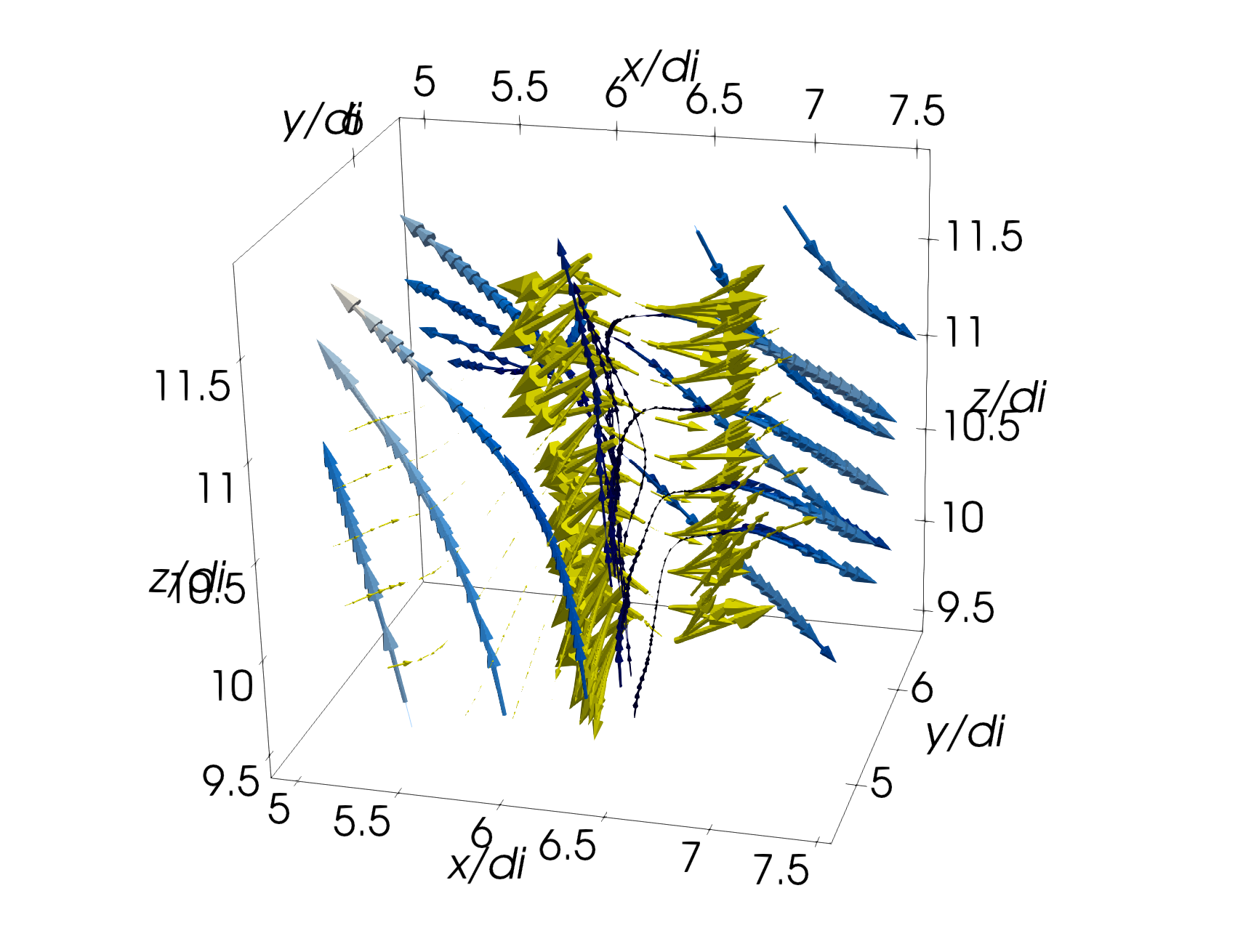}
  \includegraphics[scale=0.13]{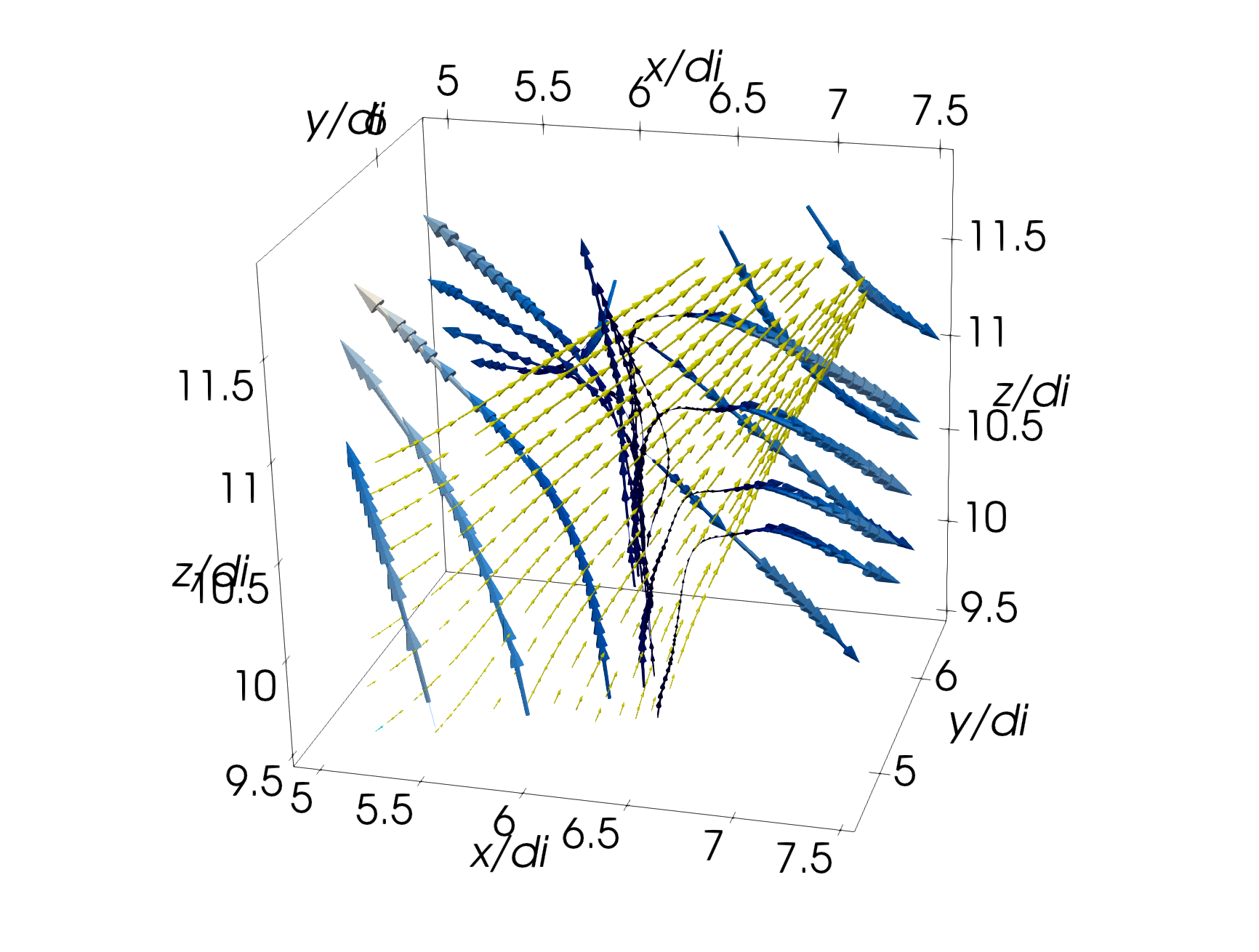}
  \caption{Reconnection event for $\beta_{\rm i} =4$, at $t = 9.2 \,\Omega_{\rm c,i}^{-1}$. The blue arrows are magnetic field lines (lower magnitude in blue, larger magnitude in white). The yellow arrows are streamlines of the velocity (normalized by the Alfvén speed) field, for the electron (left panel) and the ions (right panel).}
  \label{fig:3Dflow}
\end{figure}

\subsection{Electron-only reconnection events}\label{subsec:e-rec_events}

We now focus on the properties of electron-only reconnection events, corresponding to cases of reconnection where no ion outflows are observed. Figure \ref{fig:3Dflow} shows a 3D visualization of a reconnection event in a sub-domain of the simulation with $\beta_{\rm i}=4$. The white and blue arrows on both panels are magnetic field lines. The left panel shows streamlines of the electron velocity field, while the right panel shows those of ions at the same location. A first observation is that the presence of electron outflows are clearly visible, whereas ions just freely stream through the entire reconnection zone without being affected by it. A second observation is that the electron outflow is predominantly oriented in the plane perpendicular to the ambient magnetic field. This last characteristic has indeed been observed in each identified electron-only reconnection event. For $\beta_{\rm i} = 0.25$, the electron outflow extends over a distance $\ell_{\rm outf} \sim 0.7 d_{\rm i}$. In the $\beta_{\rm i} = 4$ case, the electron outflow is much more extended, with $\ell_{\rm outf} \sim 1.5 d_{\rm i}$.
In the $\beta_{\rm i} = 4$ case, we also notice that the decoupling between ion and electron velocities exists also outside the reconnection zone due to the generally less responsive ions over the entire sub-domain (since we are looking at scales below their Larmor radius $\ell/\rho_{\rm i}\lesssim 1$). In some cases, the outflow can be very asymmetric and predominantly in a single direction.

For each $\beta_{\rm i}$, we look at different quantities using one-dimensional cuts passing through a reconnecting kinetic-scale current sheet (i.e., a current sheet whose sizes are that of a typical site where electron-only reconnection is expected to take place).
Similarly to what is done in \citet{CalifanoFRP2020}, these cuts are meant to represent the trajectories of two virtual spacecrafts passing through these kinetic-scale current sheets. Looking at these virtual trajectories, hereafter dubbed $C_1$ and $C_2$, provides a helpful comparison with what is observed from in-situ measurements of electron-only reconnection \citep[e.g.,][]{Pha18}. For each simulation, we present two-dimensional contours of the non-ideal part of the electric field $|\boldsymbol{E}'|$, of the current density $J_\|$, and of the difference between the perpendicular ion and electron velocities $|\boldsymbol{u}_{\rm e, \perp} - \boldsymbol{u}_{\rm i , \perp}|$ in a sub-domain encompassing the selected reconnection region. Additionally, we display one-dimensional plots of the components $u_x$, $u_y$, $u_z$ for ions and electrons along the trajectories $C_1$ and $C_2$ to which we add two vertical black lines representing the boundaries of the identified reconnection region. 

\begin{figure}[t]
  \centering
  \includegraphics[scale=0.5]{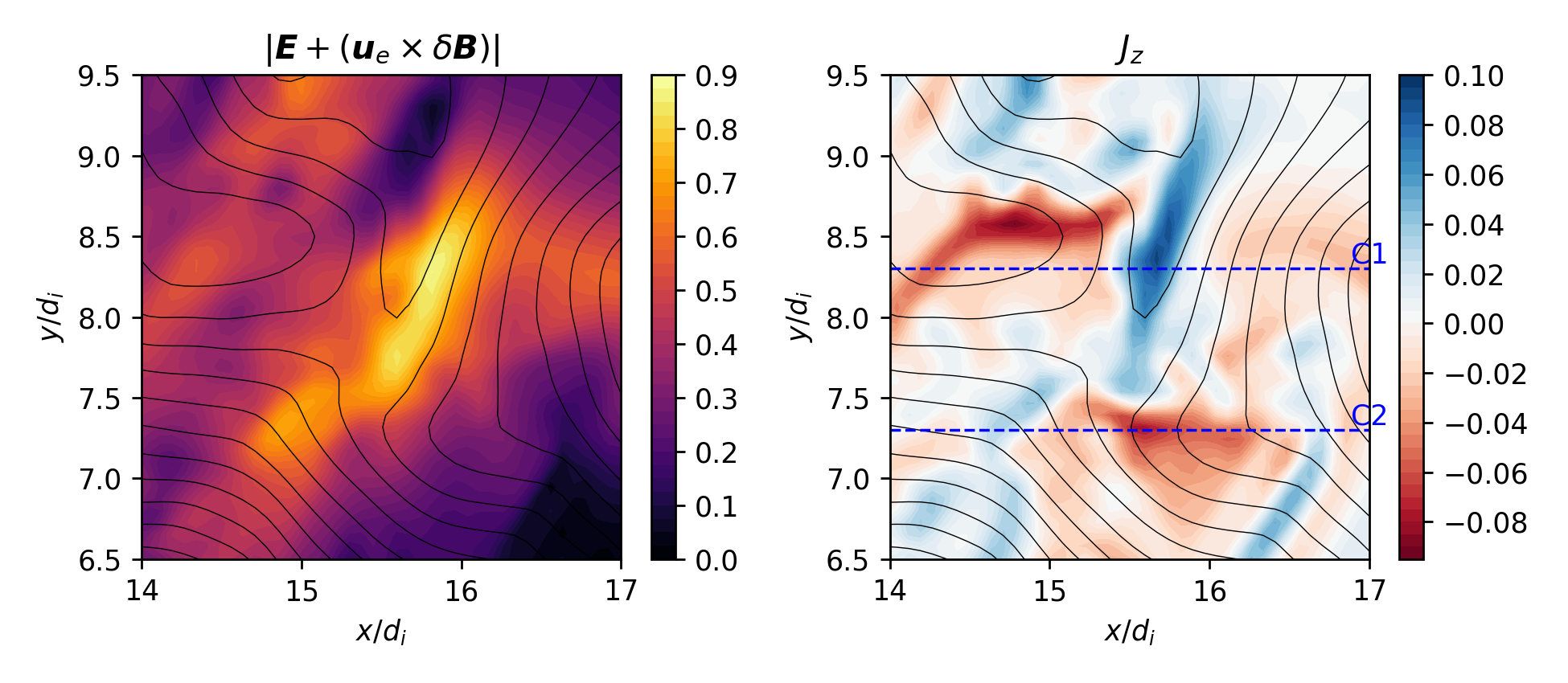}\includegraphics[scale=0.5]{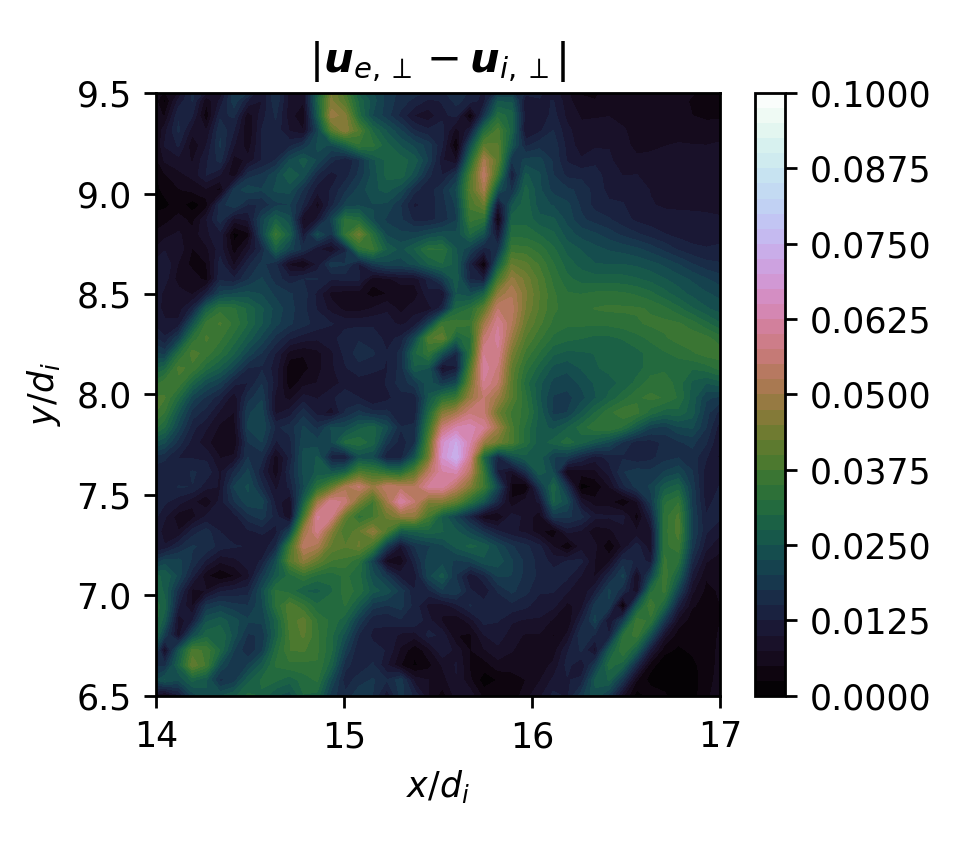}\\
  \includegraphics[width=0.4\textwidth]{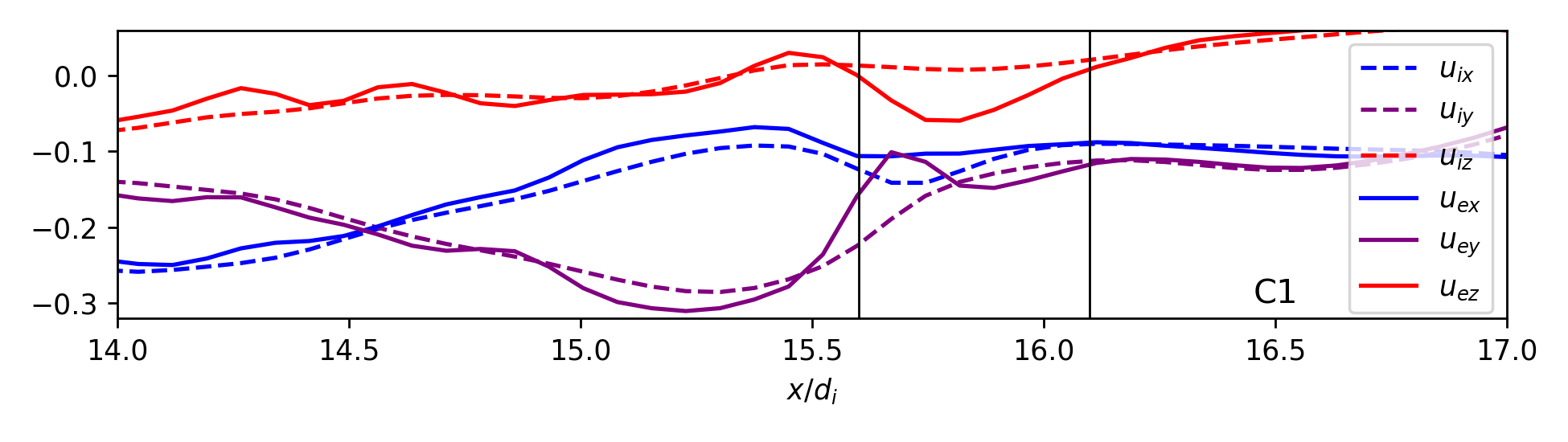}
  \includegraphics[width=0.4\textwidth]{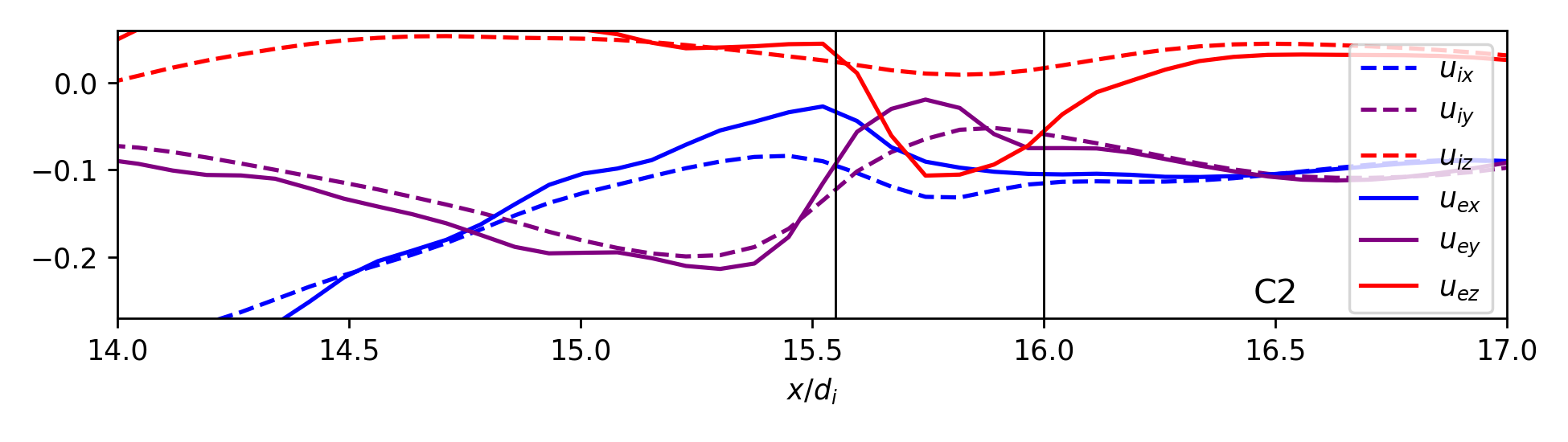}
  \caption{Various quantities at a reconnection site, in the simulation with $\beta_{\rm i} = 0.25$. Top row, from left to right: magnitude of the electric field $|\boldsymbol{E}'|$, current density $J_\|$,  absolute difference between perpendicular ion and electron velocities $|\boldsymbol{u}_{\rm e, \perp} - \boldsymbol{u}_{\rm i , \perp}|$. Bottom row, left and right: Ion and electron velocities ($u_x$, $u_y$, $u_z$) along the paths $C_1$ and $C_2$ respectively.}
  \label{fig:trabeta=0.25}
\end{figure}

Figure \ref{fig:trabeta=0.25} illustrates the case $\beta_{\rm i} = 0.25$. In this case, the absolute values of the species' velocities are overall larger than in the $\beta_{\rm i} = 1$ and $\beta_{\rm i} = 4$ cases (cf. also the spectra in Figure~\ref{fig:Spectra_ueperp_uiperp}; we recall that these flow values are in units of the background Alfv\'en speed, $v_{A,0}=B_0/\sqrt{4\pi m_{\rm i} n_0}$). In the reconnection region, there is only a slight decoupling between the species along the background magnetic field direction, with a difference of $\Delta u_z \sim 0.12$ between the ion and electron velocity. The decoupling for the $x$ and $y$ components is less important, with $\Delta u_x \sim \Delta u_y \sim 0.06$.
Nevertheless, kinetic-scale current sheets in this low-$\beta_{\rm i}$ regime seem to retain a certain coupling with the ion dynamics.

\begin{figure}[t]
  \centering
  \includegraphics[scale=0.5]{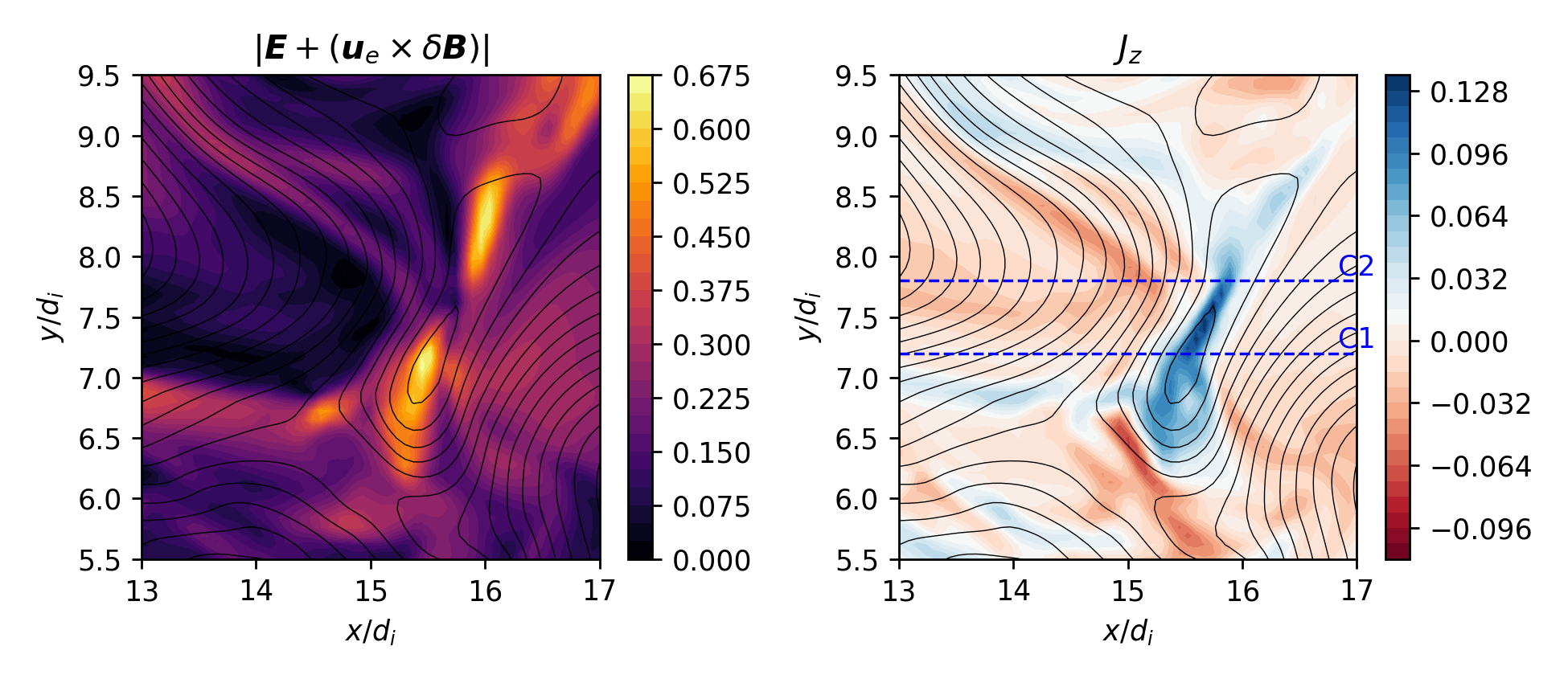}\includegraphics[scale=0.5]{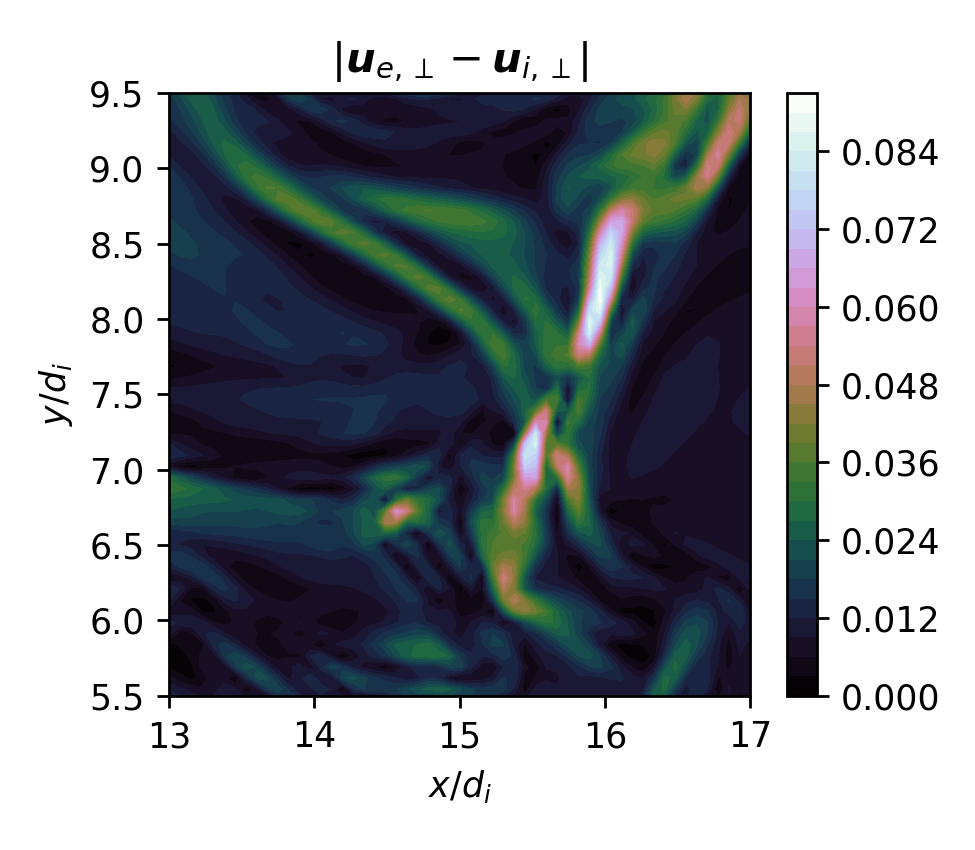}\\
  \includegraphics[width=0.4\textwidth]{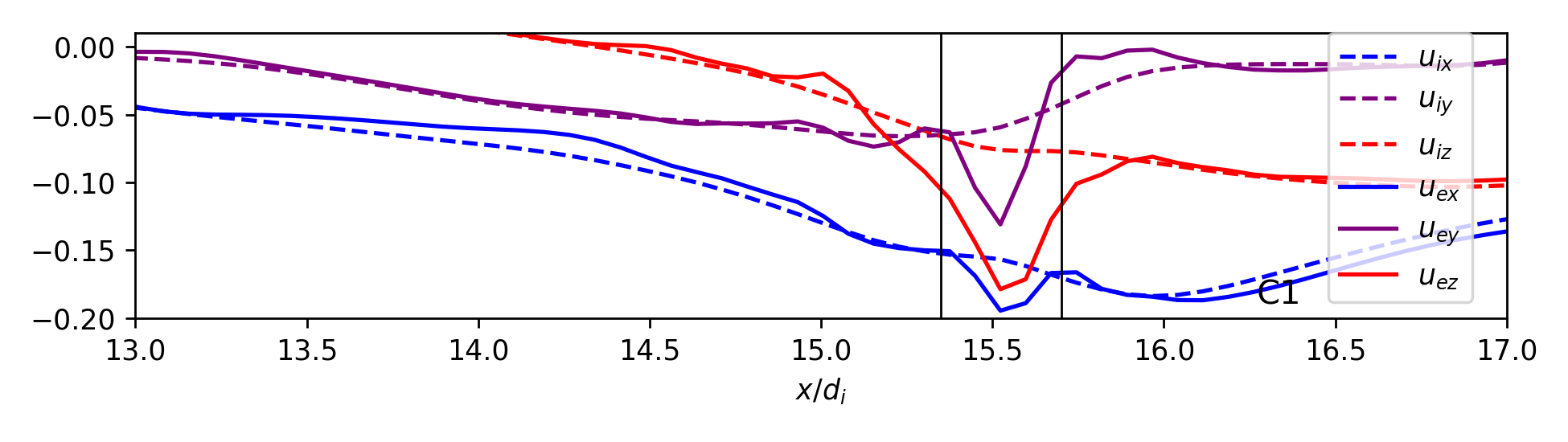}
  \includegraphics[width=0.4\textwidth]{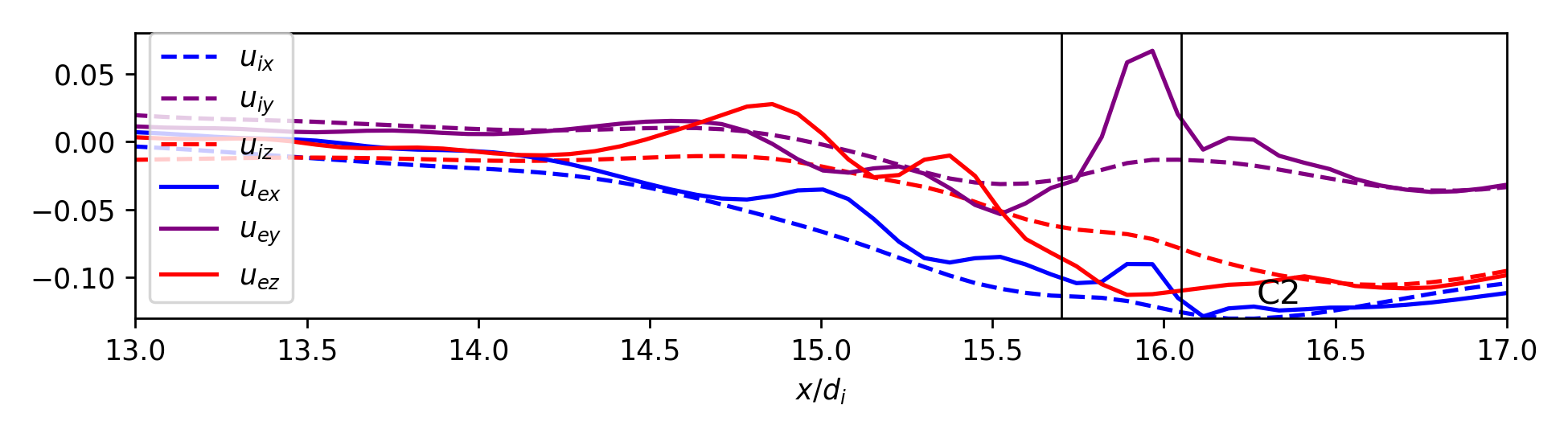}  
  \caption{Same quantities as in Figure \ref{fig:trabeta=0.25}, for the simulation with $\beta_{\rm i} = 1$.}
  \label{fig:trabeta=1}
\end{figure}

\begin{figure}[t]
  \centering
  \includegraphics[scale=0.5]{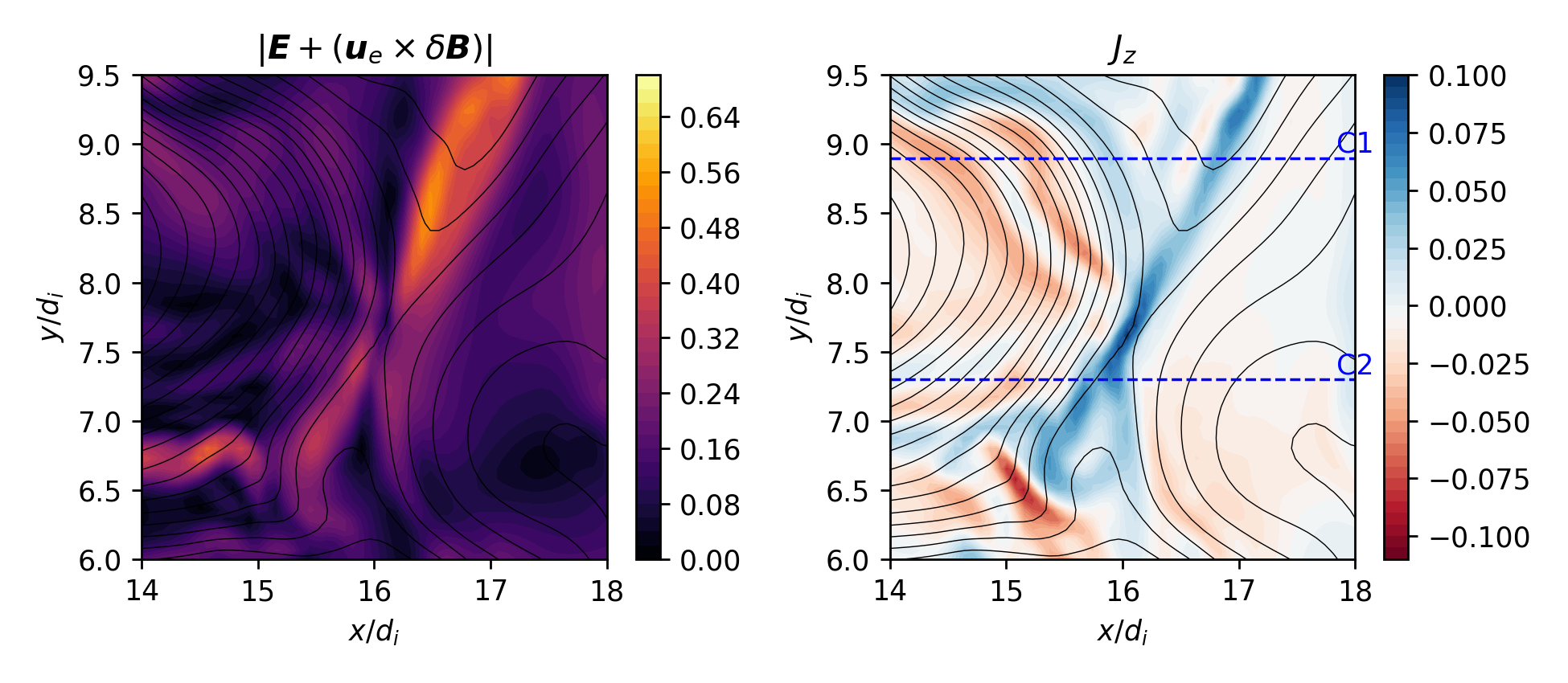}\includegraphics[scale=0.5]{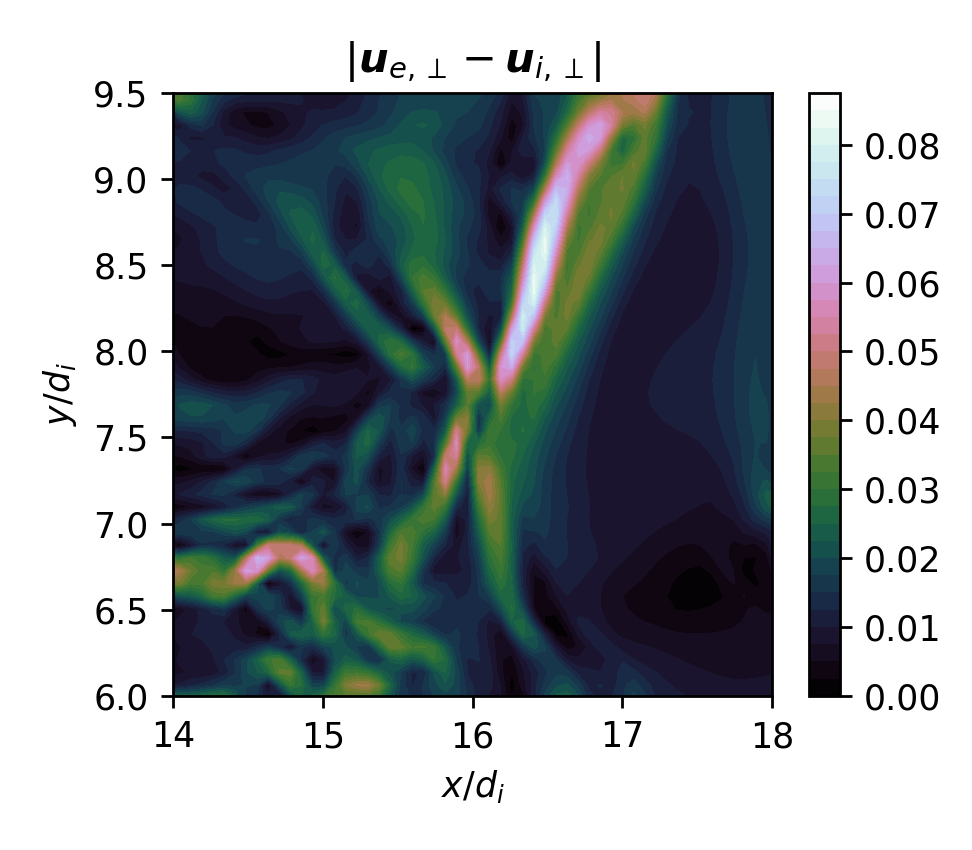}\\
  \includegraphics[width=0.4\textwidth]{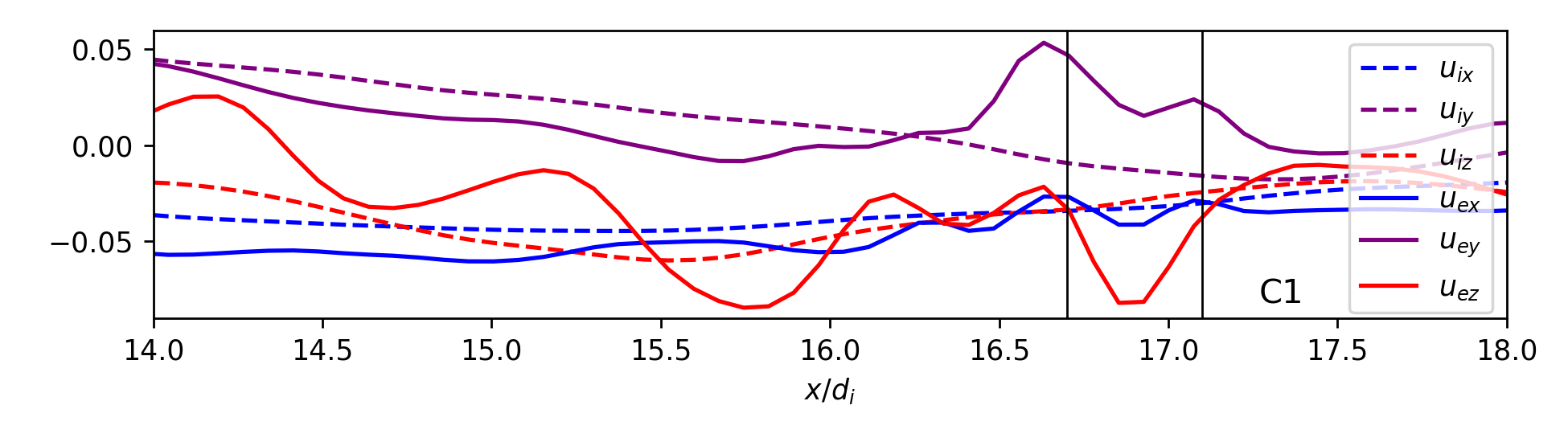}
  \includegraphics[width=0.4\textwidth]{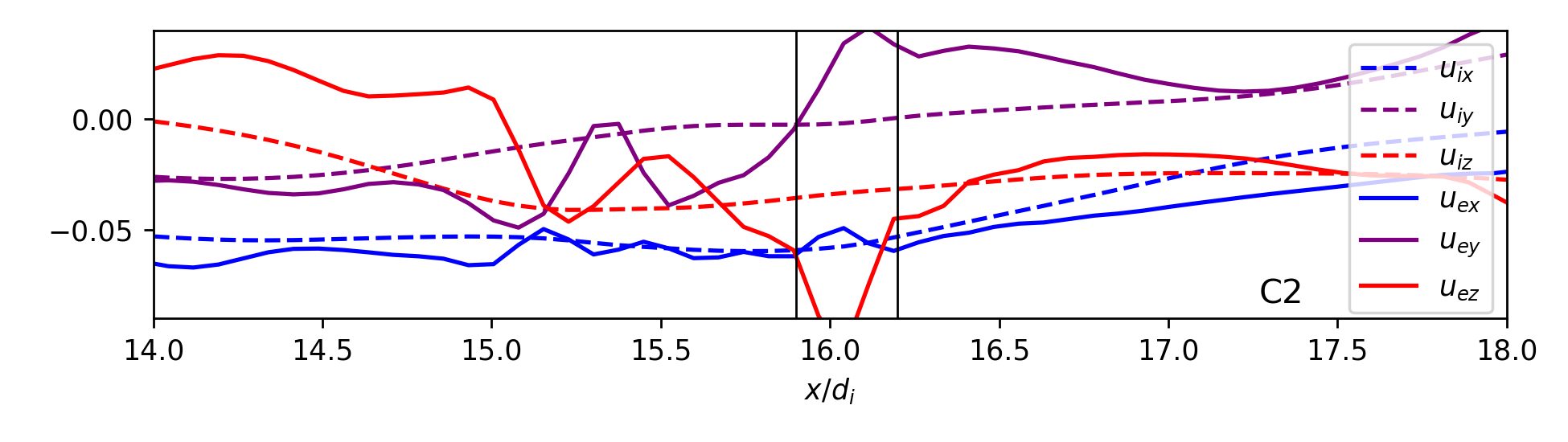}  
  \caption{ Same quantities as in Figure \ref{fig:trabeta=0.25}, for the simulation with $\beta_{\rm i} = 4$.}
  \label{fig:trabeta=4}
\end{figure}

Figure \ref{fig:trabeta=1} shows the case with $\beta_{\rm i}=1$. The decoupling between the $u_x$ and $u_y$ components of both species in the reconnection region is more pronounced, with $\Delta u_x \sim \Delta u_y \sim 0.07$ and a clear acceleration of electrons, symptomatic of electron-only reconnection is visible. Outside, yet  close to, the reconnection region, there are also small fluctuations of the electron velocities. However, as one moves away from the current sheet, the velocities of both species become nearly identical. 
For $\beta_{\rm i}=4$ (Figure \ref{fig:trabeta=4}), both species appear to be slower overall, but there is a clear decoupling of the ions and electrons throughout the entire domain, with electrons generally being faster.  In particular, the ion flow is clearly not affected by the presence of the reconnection region. The velocity difference in the reconnection zone is comparable to the previous cases, with $\Delta u_x \sim \Delta u_y \sim \Delta u_z \sim 0.05$.

Overall, we observe a decoupling between ions and electrons at sub-ion scales -- and thus the occurrence of electron-only reconnection -- that is more prominent at larger $\beta_{\rm i}$. Since $\lambda_{\rm inj}/d_{\rm i}$ is fixed in our setup ($\lambda_{\rm inj}$ being the typical injection scale of turbulent fluctuations), our simulations support the idea that the occurrence of the electron-only reconnection regime is more sensitive to the separation between the injection scale $\lambda_{\rm inj}$ and the ion Larmor radius $\rho_{\rm i}$, rather than between $\lambda_{\rm inj}$ and the ion inertial length $d_{\rm i}$.

\subsection{Ion turbulent heating}\label{subsec:ion_heating}

Let us now focus on the turbulent heating of ions, i.e., how the energy of turbulent electromagnetic fluctuations (characterized in Section~\ref{subsec:spectra}) is converted into ion thermal energy along their cascade towards smaller scales.

We start by analyzing the change in ion temperature parallel and perpendicular to the {\it local} magnetic-field direction $\boldsymbol{b}=\boldsymbol{B}/|\boldsymbol{B}|$, i.e., $T_{\rm i,\|}=(\boldsymbol{\Pi}_{\rm i}:\boldsymbol{b}\boldsymbol{b})/n=(\Pi_{{\rm i},ij}b_ib_j)/n$ and $T_{\rm i,\perp}=(\boldsymbol{\Pi}_{\rm i}:\boldsymbol{\sigma})/n=(\Pi_{{\rm i},ij}\sigma_{ij})/n$, where $\boldsymbol{\Pi}_{\rm i}$ is the ion pressure tensor and $\sigma_{ij}=(\delta_{ij}-b_ib_j)/2$ is the projector onto the plane perpendicular to $\boldsymbol{B}$ ($\delta_{ij}$ is the Kronecker delta). 
This information will provide helpful in supporting the interpretation of the ion-heating mechanisms that are likely operating in our HVM simulations at different $\beta_{\rm i}$.

\begin{figure}[t]
  \centering
\includegraphics[width=0.495\textwidth]{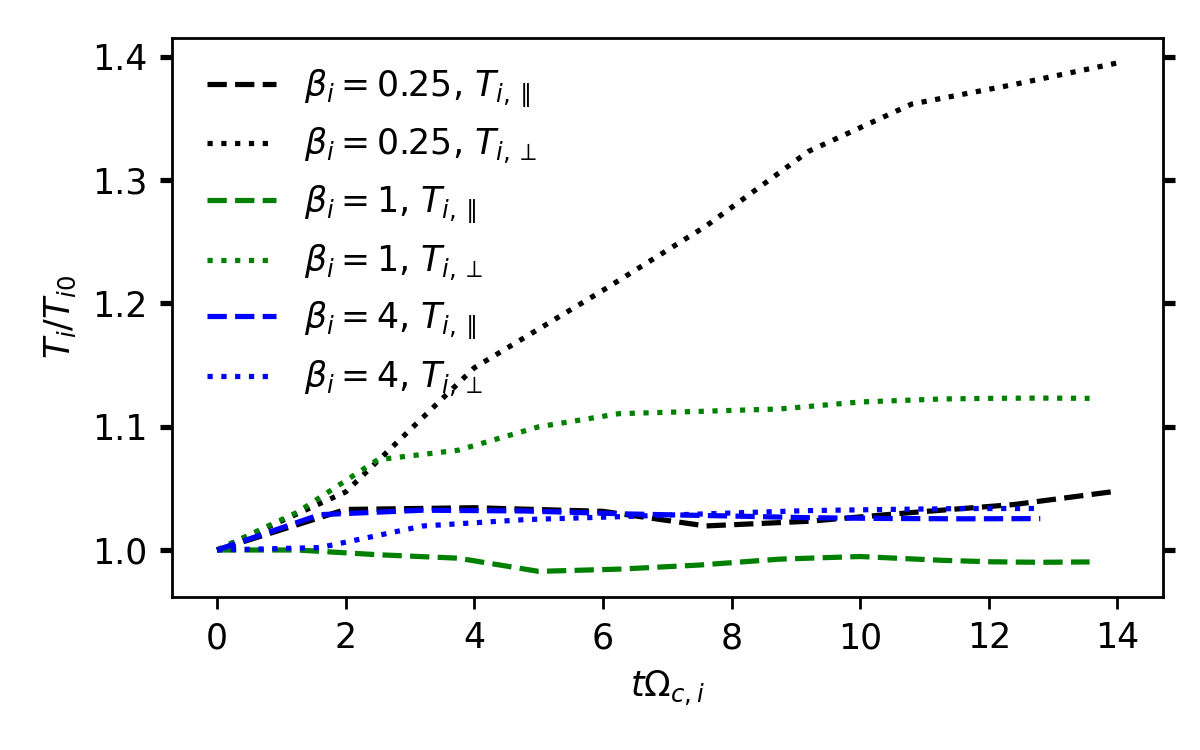}
\includegraphics[width=0.495\textwidth]{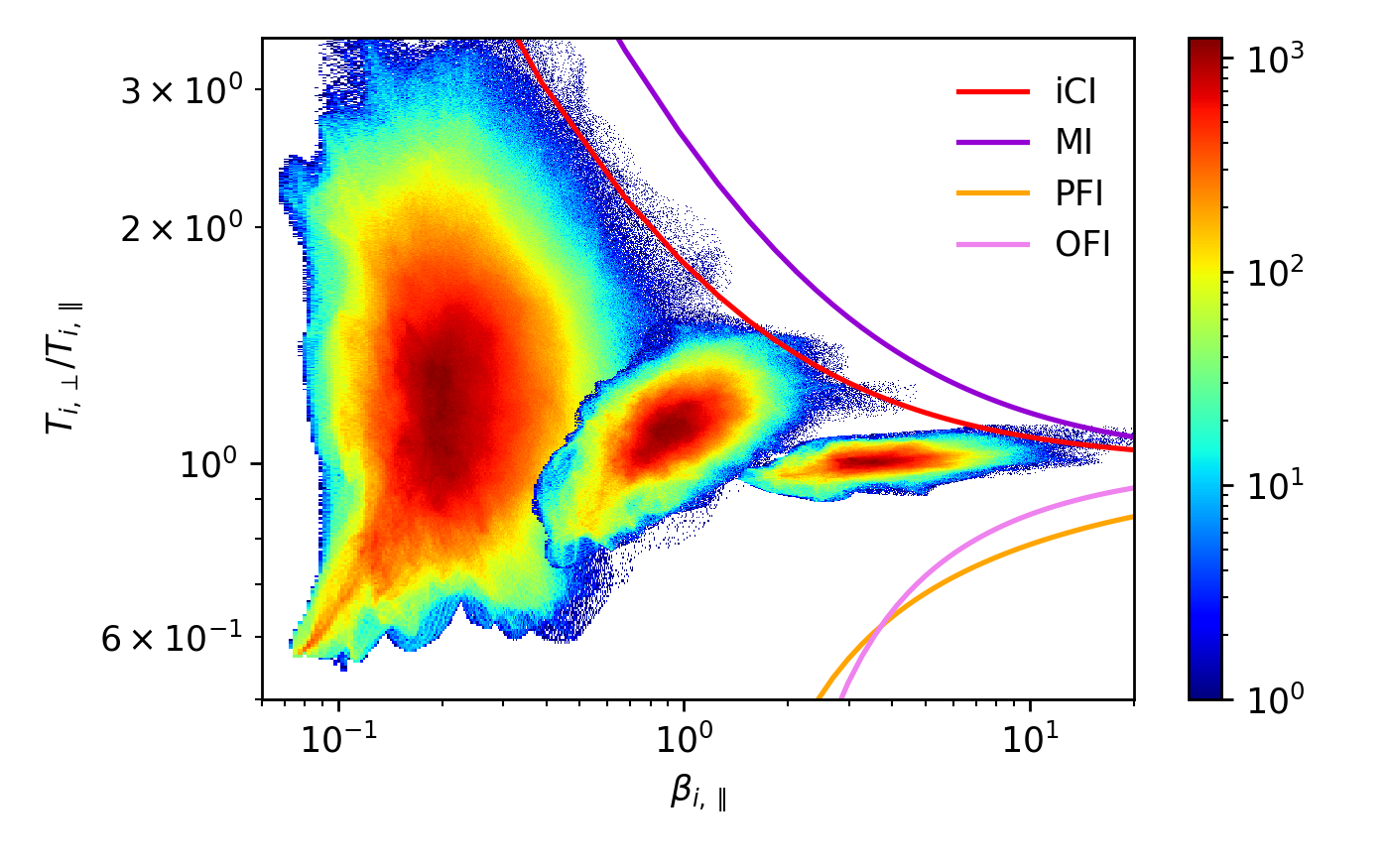}
  \caption{Left: time evolution of parallel and perpendicular ion temperatures, $T_{\rm i,\perp}/T_{\rm i0}$ (dashed lines) and $T_{\rm i,\|}/T_{\rm i0}$ (dotted lines), where $T_{\rm i0}$ is the initial (isotropic) temperature of the ions, in our HVM simulations with different initial $\beta_{\rm i}$ (see legend). Right: Histogram of the ($\beta_{\rm i,\|}$, $T_{\rm i,\perp}/T_{\rm i,\|}$) values occurring in the simulation domain and over a time interval around the peak turbulent activity for HVM runs with $\beta_{\rm i}=0.25$ (left distribution), $\beta_{\rm i}=1$ (center distribution), and $\beta_{\rm i}=4$ (right distribution); different histograms have been normalized so that their maximum values coincide. Solid lines represent the thresholds with a maximum growth rate $\gamma_{\rm max}=10^{-2}\Omega_{\rm c,i}^{-1}$ for various proton-anisotropy driven instabilities~\citep[from Fig.5 in][]{FinelliAA2021}: mirror instability (MI, dark violet line), ion-cyclotron instability (iCI, red line), parallel firehose instability (PFI, orange line), and oblique firehose instability (OFI, pink line).}
   \label{fig:Tprp_Tpar_Aniso-beta}
\end{figure}

In the left plot of Figure~\ref{fig:Tprp_Tpar_Aniso-beta}, we show the time evolution of the box-averaged $T_{\rm i,\perp}$ and $T_{\rm i,\|}$ (normalized to the initial isotropic ion temperature $T_{\rm i0}$) in our HVM simulations with initial ion beta $\beta_{\rm i}=0.25$ (black curves), $\beta_{\rm i}=1$ (green curves), $\beta_{\rm i}=4$ (blue curves). 
One can notice that the temperature increase is anisotropic, preferentially heating the ions in the direction perpendicular to the magnetic field (i.e. $T_{\rm i,\perp}>T_{\rm i,\|}$), and that this tendency is more pronounced as the $\beta_{\rm i}$ decreases.
This trend is confirmed also by local estimates, as shown in the right plot of Figure~\ref{fig:Tprp_Tpar_Aniso-beta}, where we report a cumulative distribution of the values in a parameter space described by ion-temperature anisotropy, $T_{\rm i,\perp}/T_{\rm i,\|}$, and parallel ion beta, $\beta_{\rm i,\|}=8\pi n T_{\rm i,\|}/B^2$, that are occurring in our HVM simulations for the interval of time around the turbulent peak activity: $\beta_{\rm i}=0.25$ (left distribution), $\beta_{\rm i}=1$ (center distribution), and $\beta_{\rm i}=4$ (right distribution).
In that anisotropy-beta plane, we also draw solid lines representing the same thresholds for proton-anisotropy driven instabilities (with a maximum growth rate of $\gamma_{\rm max}=10^{-2}\Omega_{\rm c,i}^{-1}$) that were used in \citet{FinelliAA2021}. In particular, the ion-cyclotron instability (iCI) threshold is from \citet{LazarPoedtsMNRAS2014}, the mirror instability (MI) curve is from \citet{MarucaAPJ2012}, while the thresholds for parallel and oblique firehose instabilities are from \citet{AstfalkJenkoJGRA2016}.
This type of representation for the plasma distribution is widely used in solar-wind turbulence studies~\citep[e.g.,][]{HellingerGRL2006,MatteiniGRL2007,BalePRL2009,ChenAPJ2016,MarucaAPJ2018,BandyopadhyayPOP2022}.
One can appreciate that in $T_{\rm i,\perp}/T_{\rm i,\|}$ the plasma distribution exhibit a larger spread as the $\beta_{\rm i}$ decreases: this is likely the result of anisotropy driven instabilities (and especially of the iCI, for our setup) in bounding these distributions to $T_{\rm i,\perp}/T_{\rm i,\|}$ values that are within a marginally stable region (which indeed gets narrower as $\beta_{\rm i,\|}$ increases).
The associated spread in $\beta_{\rm i,\|}$ is likely related to the heating (or cooling) parallel to the magnetic field, as well as to local density fluctuations and variation of magnetic-field strength (mainly associated to $\delta B_\|$ fluctuations). In particular, despite the relatively narrow anisotropy spread, even the simulation with initial $\beta_{\rm i}=4$ exhibit a large spread in $\beta_{\rm i,\|}$. This, combined with the fact that the parallel heating is comparable to the perpendicular one for the $\beta_{\rm i}=4$ simulation (Figure~\ref{fig:Tprp_Tpar_Aniso-beta}, left panel) and that this regime shows significant magnetic compressibility (cf. Figure~\ref{fig:MagCompr}, left panel), can be interpreted as a signature of transit-time damping (TTD) mediated by the magnetic-mirror force~\citep{BarnesPOF1966} being a relevant process in our $\beta_{\rm i}=4$ run.

In order to estimate the turbulent heating rate for the ions, we use the following procedure. Since the initial condition consists of only magnetic-field fluctuations (cf Section~\ref{subsec:sim_setup}), this represent the injected fluctuation energy that will cascade towards smaller scales. Therefore, we compare the decrease in magnetic energy, $|\Delta E_{\rm mag}|=|E_{\rm mag}(t)-E_{\rm mag}(0)|$, where $E_{\rm mag}=\langle \delta B^2\rangle/2$, with the simultaneous increase in ion thermal energy, $\Delta E_{\rm th,i}=E_{\rm th,i}(t)-E_{\rm th,i}(0)$, where $E_{\rm th,i}=\langle{\rm tr}[\Pi_{\rm i}]\rangle/2$ and ${\rm tr}[\Pi_{\rm i}]$ is the trace of the ion pressure tensor. The ratio between these two quantities is then used as a proxy for the ratio between the ion-heating rate $Q_{\rm i}$ and the cascading rate $\varepsilon$, namely, $\Delta E_{\rm th,i}/|\Delta E_{\rm mag}|\approx Q_{\rm i}/\varepsilon$. Moreover, we use this proxy to also provide an estimate of the ion-to-electron heating ratio, i.e., by computing the ratio $Q_{\rm i}/Q_{\rm e}\simeq Q_{\rm i}/(\varepsilon-Q_{\rm i})\approx(\Delta E_{\rm th,i}/|\Delta E_{\rm mag}|)[1-(\Delta E_{\rm th,i}/|\Delta E_{\rm mag}|)]^{-1}$.
As done in previous hybrid-kinetic studies~\citep[e.g.,][]{ArzamasskiyAPJ2019,CerriAPJ2021,Arz23,SquireNatAs2022}, by using this approximation we implicitly assume that while the injected energy (consisting of only $\delta B$ fluctuations, in our setup) cascades towards smaller scale (represented by the decrease in $E_{\rm mag}$), a fraction of it will be converted into ion heating across and below the ion scales (consisting in the increase of $E_{\rm th,i}$), while all the rest will be eventually dissipated into electron heating only at the the electron scales (represented by all the energy that is not accounted for in $E_{\rm th,i}$, which is eventually dissipated by numerical filters in our hybrid-kinetic simulations). 
This approach to estimate the ion heating is intimately related to the hybrid-kinetic approximation itself, i.e., to the assumption that electron kinetic effects (and thus significant electron heating) can be neglected at ion scales, assumed to be much larger than the electron gyroradius scale.
We also caution that, in general, the crude heating-rate estimate outlined above would neglect some residual energy that can be left in the bulk flows of the two species~\citep[e.g., see Fig.1 in][]{CerriCamporealePOP2020}. However, these bulk-flow contributions are negligible in our simulations, since the amount of energy channeled into ion flows is strongly suppressed at sub-ion scales (especially for the $\beta_{\rm i}\geq1$ cases; cf. spectra in Figure~\ref{fig:Spectra_ueperp_uiperp}) and, despite the fact that our setup favours a large amount of small-scale jets associated to electron-only reconnection, the electron-flow energy involves a factor $m_{\rm e}/m_{\rm i}\ll1$ compared to the ion-flow energy, thus making this channel also negligible.

\begin{figure}[t]
  \centering
  \includegraphics[width=0.495\textwidth]{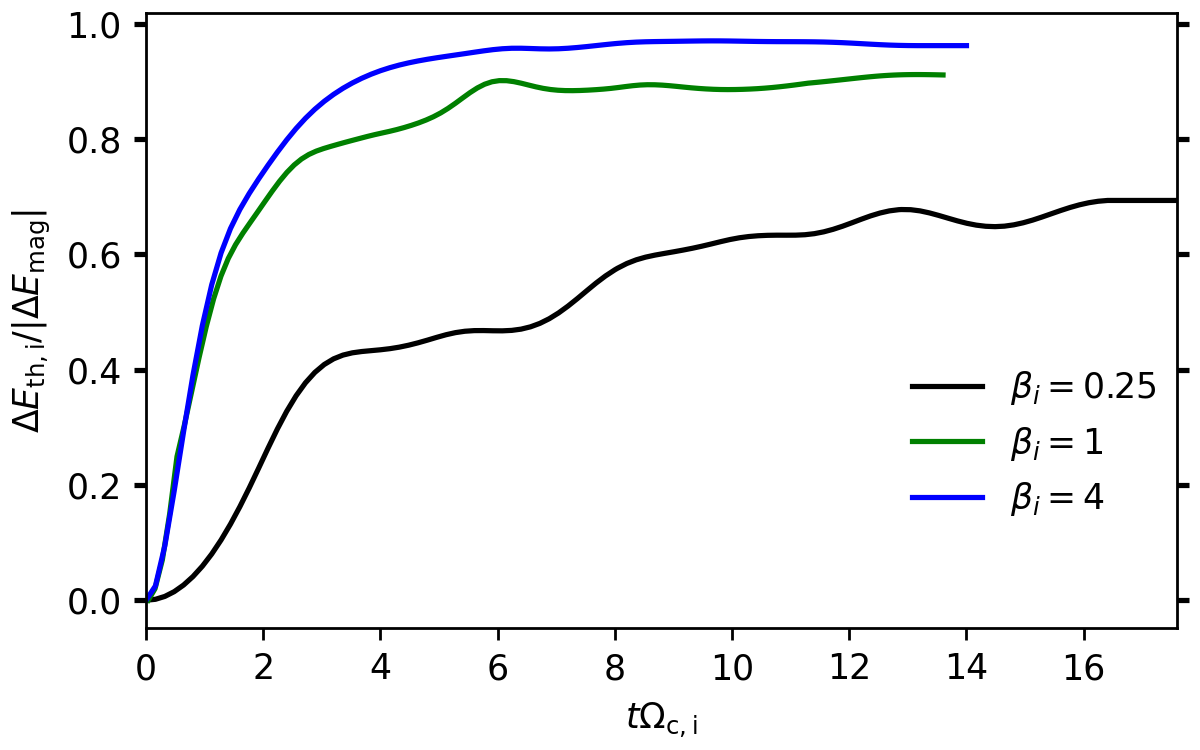}
  \includegraphics[width=0.495\textwidth]{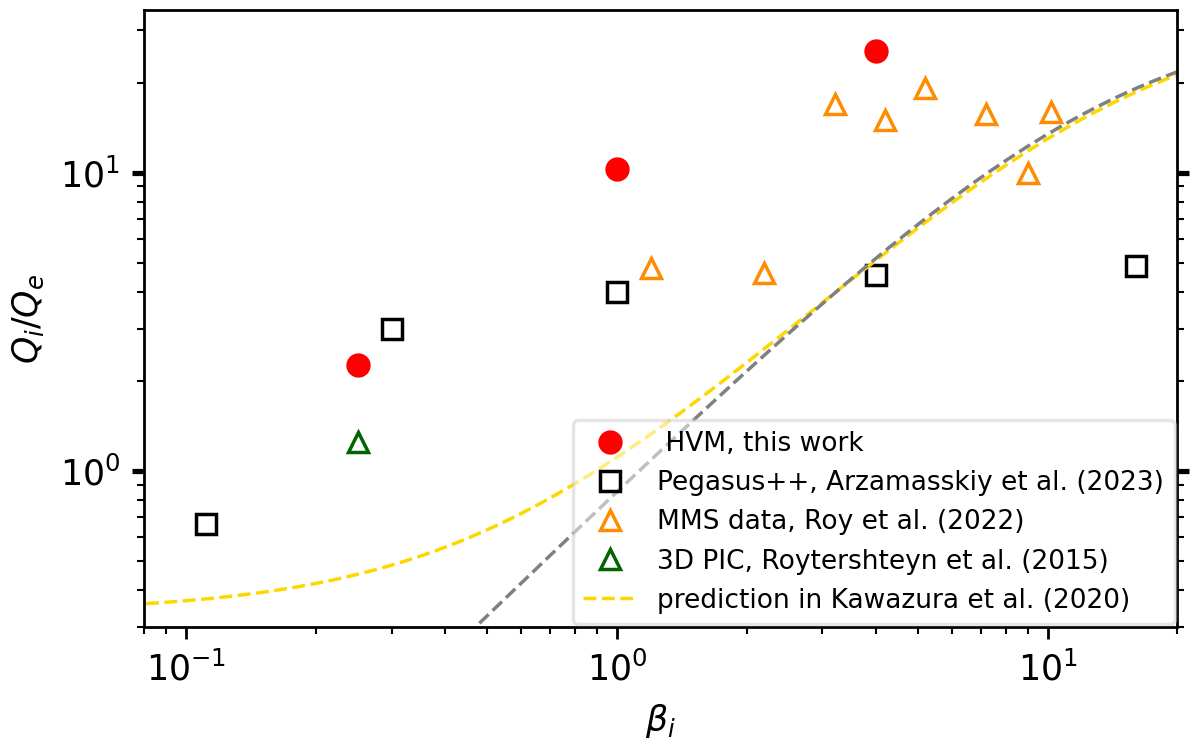}
  \caption{Left: time evolution of $\Delta E_{\rm th,i}/|\Delta E_{\rm mag}|\approx Q_{\rm{i}}/\varepsilon$ in our simulations with different initial $\beta_{\rm i}$ regimes (see legend). Right: Inferred ion-to-electron heating ratio $Q_{\rm i}/Q_{\rm e}$ versus the initial ion beta $\beta_{\rm i}$ from our HVM simulations with injection of compressive fluctuations ($\delta B_\|\neq0$) at $k_{\rm inj}d_{\rm i}\sim1$ (red circles). For comparison, we also report the heating ratio obtained with Pegasus++ simulations of Alfv\'enic turbulence ($\delta B_\|=0$) continuously driven at scales $k_{\rm\perp,inj}\rho_{\rm i}\sim0.1$--$0.2$~\citep[black squares; from Fig.~15(a) of][]{Arz23}, the ratio $Q_{\rm i}/Q_{\rm e}$ inferred from a 3D PIC simulation of freely decaying compressive fluctuations with $\delta B/B\sim1$ injected up to $k_{\rm inj}d_{\rm i}\sim1$ in a $\beta_{\rm i}=\beta_{\rm e}=0.25$ plasma (green triangle; original simulation from \citet{RoytershteynRSPTA2015}, corresponding heating ratio reported in Table~1 of \citet{Roy22}), as well as the ion-to-electron heating measured by MMS in the Earth's magnetosheath~\citep[orange triangles; from Fig.~6 of][]{Roy22}. Two curves obtained with the fitting formula for $Q_{\rm i}/Q_{\rm e}$ based on hybrid-gyrokinetic simulations~\cite[eq.(14) of][]{KawazuraPRX2020} are also plotted: a curve adopting parameters corresponding to our setup, i.e., $T_{\rm e}/T_{\rm i}=\beta_{\rm e}/\beta_{\rm i}=0.1/\beta_{\rm i}$ and an estimated ratio of compressible-to-incompressible injection power $P_{\rm compr}/P_{\rm AW}\approx1/3$ (yellow dashed line; to be compared with red circles), as well as a curve for the Pegasus++ setup, i.e., $T_{\rm e}/T_{\rm i}=1$ and $P_{\rm compr}/P_{\rm AW}=0$ (grey dashed line; to be compared with black squares).}
   \label{fig:Qi_Tprp_Tpar}
\end{figure}

The time evolution of the ratio $\Delta E_{\rm th,i}/|\Delta E_{\rm mag}|$ is shown in Figure~\ref{fig:Qi_Tprp_Tpar} (left panel), and develops a plateau when the simulations achieve the peak turbulent activity, denoting a saturation of the ion turbulent heating; the ion heating $Q_{\rm i}/\varepsilon$ for a given simulation is estimated by averaging $\Delta E_{\rm th,i}/|\Delta E_{\rm mag}|$ over this quasi-steady phase. From these curves it is evident that a larger fraction of the cascading magnetic energy is converted into ion heating as $\beta_{\rm i}$ increases, namely from $Q_{\rm i}/\varepsilon\simeq 69\%$ at $\beta_{\rm i}=0.25$, to $Q_{\rm i}/\varepsilon\simeq 91\%$ at $\beta_{\rm i}=1$ and $Q_{\rm i}/\varepsilon\simeq 96\%$ at $\beta_{\rm i}=4$.
The percentage of ion turbulent heating in our $\beta_{\rm i}=0.25$ case is consistent with the trend obtained by previous hybrid-PIC simulations at low $\beta_{\rm{i}}$~\citep[obtaining $Q_{\rm i}/\varepsilon\approx75\%$ at $\beta_{\rm i}=0.3$ and $Q_{\rm i}/\varepsilon\approx40\%$ at $\beta_{\rm i}\approx0.1$; see][]{ArzamasskiyAPJ2019,CerriAPJ2021}. On the other hand, we observe a larger ion heating in our $\beta_{\rm i}\geq1$ cases, when compared to previous results obtained with hybrid-PIC simulations in the same $\beta_{\rm i}$ regimes~\citep[obtaining $Q_{\rm i}/\varepsilon\approx80\%$ at $\beta_{\rm i}=1$ and $Q_{\rm i}/\varepsilon\approx85\%$ at $\beta_{\rm i}=4$; see][]{ArzamasskiyAPJ2019,Arz23}. 
The trend of ion-to-electron heating ratio $Q_{\rm i}/Q_{\rm e}$ versus $\beta_{\rm i}$ that has been estimated with our $3$D hybrid-Vlasov simulations of freely decaying turbulence is reported in the right panel of Figure~\ref{fig:Qi_Tprp_Tpar} (red circles), where it is compared with some previous results from $3$D simulations and in-situ observations that are reported in the literature.
In particular, we include the heating ratio estimated via $3$D hybrid-PIC simulations of continuously driven Alfv\'enic turbulence~\citep[black squares; from][]{ArzamasskiyAPJ2019,CerriAPJ2021,Arz23}, as well as the ratio inferred from a 3D full-PIC simulation of freely decaying turbulence with large-amplitude injection of compressive fluctuations at ion scales~\citep[gree triangle; from][]{RoytershteynRSPTA2015,Roy22}.
The different behavior between HVM and Pegasus++ results at $\beta_{\rm i}\geq1$, is mainly related to the difference in the injection scales and, to a certain extent, to a different nature and amplitude of fluctuations that are injected in these two set of simulations.
In fact, with respect to our setup, the hybrid-PIC simulations continuously drive smaller-amplitude and more anisotropic, purely Alfv\'enic fluctuations at scales larger than the ion gyroradius scale ({\it viz.}, transverse magnetic and velocity fluctuations, $\delta B_\|=\delta u_{\rm i,\|}=0$, are driven at scales up to $k_{\rm \perp,inj}^{\rm max}\rho_{\rm i}\sim0.1$--$0.2$ and with $k_{\rm\|,inj}/k_{\rm\perp,inj}\sim0.1$--$0.5$, achieving $\delta B_{\rm\perp,rms}/B_0\sim\delta u_{\rm i\perp,rms}/v_{\rm A}\sim0.1$--$0.5$ in the quasi-stationary state; some of these parameters vary significantly among different $\beta_{\rm i}$ regimes).
On the other hand, our HVM simulations initialize large-amplitude, quasi-isotropic, compressive magnetic-field fluctuations that freely decay into a fully developed turbulent state ({\it viz.}, only magnetic-field fluctuations with $\delta B_\|\neq0$ are initialized with $\delta B_{\rm rms,0}/B_0\sim0.5$ and at scales up to $k_{\rm inj}^{\rm max}d_{\rm i}\sim1$, then decaying to a level $\delta B_{\rm rms}/B_0\sim0.3$ around peak turbulent activity).
In that context, the heating ratio that we obtain at $\beta_{\rm i}=0.25$ seems to align well with the result obtained by a 3D full-PIC simulation with the same ion beta in which compressive fluctuations of amplitude $\delta B/B\sim1$ were injected up to $k_{\rm inj}^{\rm max}d_{\rm i}\sim1$~\citep{RoytershteynRSPTA2015,Roy22}. The slightly lower value of $Q_{\rm i}/Q_{\rm e}$ observed in the PIC run is likely due to a combination of two factors. One is the fact that the PIC simulation adopts an electron beta of $\beta_{\rm e}=0.25$, which is larger than the $\beta_{\rm e}=0.1$ used in our setup, and the other is that the authors employ an even smaller reduced mass ratio of $m_{\rm i}/m_{\rm e}=50$ (while we use $m_{\rm i}/m_{\rm e}=100$). These two effects combine to reduce significantly the separation of scales between $\rho_{\rm i}$ and $\rho_{\rm e}$ in the PIC simulation (namely, $\rho_{\rm i}/\rho_{\rm e}\approx7$, while $\rho_{\rm i}/\rho_{\rm e}\approx16$ in our setup), thus enabling non-negligible electron heating close to the ion scales and therefore reducing the ion-to-electron heating ratio with respect to what is estimated with our simulations.
As outlined in Section~\ref{subsec:setup_justification}, our choice of injection properties and plasma parameters is motivated by several observations in the Earth's magnetosheath, past the bow shock; and in general, it can be relevant for those space and astrophysical environments where several plasma instabilities may concur in directly injecting fluctuations' energy close to the ion scales.
To better quantify this aspect in the context of the present study, the right panel of Figure~\ref{fig:Qi_Tprp_Tpar} also includes the ion-to-electron heating ratio measured in-situ by the MMS mission within the Earth's magnetosheath~\citep[orange triangles; from][]{Roy22}. Comparing these $Q_{\rm i}/Q_{\rm e}$ measurements from MMS with our HVM results, one can see that there is a qualitative agreement between the trend of the two set of data and, in particular, a more quantitative agreement at large ion beta, $\beta_{\rm i}\sim4$ (where the HVM simulations better match the Earth's magnetosheath conditions). We also mention that, interestingly, MMS measurements of $Q_{\rm i}/Q_{\rm e}$ in the $\beta_{\rm i}\sim1$ regime seem to be in line with the heating ratio associated to purely Alfv\'enic turbulence obtained with hybrid-PIC simulations~\citep{ArzamasskiyAPJ2019}. This latter observation, together with the fact that MMS measurements agree with the results from our simulation at $\beta_{\rm i}\sim4$, potentially calls for a more detailed analysis and classification of the various MMS intervals used in \citet{Roy22}, which, at a given $\beta_{\rm i}$, may be dominated by turbulent fluctuations of different nature.
In addition to the above, we mention also the work of \citet{Roy24}, where they studied the local heating of a reconnecting current sheet, showing more electron than ion heating occurring in electron-only reconnection sites, i.e., $Q_{\rm i}<Q_{\rm e}$. The discrepancy between such heating ratio and the results obtained from our simulations and from MMS measurements~\citep{Roy22}, for which $Q_{\rm i} > Q_{\rm e}$, can be understood in terms of the simulated scales. In fact, to estimate $Q_{\rm i}/Q_{\rm e}$, \citet{Roy24}  only considered the local reconnection sites, which are electron-scale structures where the ions are almost undisturbed. As a result, their set up does not account for the ion heating occurring in the turbulent fluctuations that are populating the volume outside the very small-scale electron-only reconnection sites. Our simulations suggest that triggering electron-only reconnection in a larger-scale turbulent system requires injecting fluctuations at scales comparable to $\rho_{\rm i}$, leading to significant ion heating (e.g., stochastic heating), which eventually dominates over the very local electron heating due to electron-only reconnection.
Finally, for completeness, we also include two lines obtained with the fitting formula for $Q_{\rm i}/Q_{\rm e}$ based on hybrid-gyrokinetic simulations~\cite[gyrokinetic ions and fluid electrons; see eq.(14) of][]{KawazuraPRX2020}; the yellow dashed line (to be compared with red circles) represents the prediction obtained adopting plasma parameters corresponding to our setup, i.e., $T_{\rm e}/T_{\rm i}=\beta_{\rm e}/\beta_{\rm i}=0.1/\beta_{\rm i}$ and a ratio of compressible-to-incompressible injection power that in our simulations has been estimated to be $P_{\rm compr}/P_{\rm AW}\approx1/3$, while the grey dashed line (to be compared with black squares) represents the prediction for the heating ratio in purely Alfv\'enic turbulence pertaining to the hyrbid-PIC simulations, i.e., $T_{\rm e}/T_{\rm i}=1$ and $P_{\rm compr}/P_{\rm AW}=0$. When comparing predictions for $Q_{\rm i}/Q_{\rm e}$ based on hybrid-gyrokinetics with the corresponding results from hybrid-kinetic simulations (yellow dashed line vs red circles, and grey dashed line vs black squares), one can see that the former almost always underestimate the heating ratio by an order of magnitude or more, depending on the case; the only exception to this behaviour being the case of purely Alfv\'enic turbulence, where the predictions overestimate the simulation results by nearly an order of magnitude at high $\beta_{\rm i}$. The reason for the discrepancy between the analytical predictions and the results from hybrid-kinetic (and full-kinetic) simulations and MMS measurements is due to the fact that the model by \citet{KawazuraPRX2020} is based on the gyrokinetic ordering~\citep[see, e.g.,][]{SchekochihinAPJS2009} which only retains a limited number of ion-heating mechanisms, namely the parallel heating associated to ion-Landau damping (iLD) and transit-time damping~\citep[TTD;][]{BarnesPOF1966} as collisionless damping mechanism for the fluctuations (plus finite-collisionality heating arising from nonlinear phase mixing). Therefore, by neglecting perpendicular ion-heating mechanisms such as stochastic heating and cyclotron heating, the hybrid-gyrokinetic model will inevitably underestimate $Q_{\rm i}$ with respect to a hybrid-kinetic model (especially at $\beta_{\rm i}\lesssim1$).

To summarize, our results thus suggest that, as $\beta_{\rm i}$ increases, a larger fraction of turbulent energy is channeled into ion heating when fluctuations are excited close to the ion scales---and that this happens despite a sub-ion-scale dynamics that is better described by the EMHD regime at larger $\beta_{\rm i}$, and the consequent prevalence of electron-only reconnection events.
Moreover, such ion heating is strongly anisotropic with respect to the (local) direction of the magnetic field at $\beta_{\rm i}\leq1$ (occurring mostly in the perpendicular direction, $T_{\rm i,\perp} \gg T_{\rm i,\|}$), while it is only moderately anisotropic for $\beta_{\rm i}=4$ ({\it viz.}, $T_{\rm i,\perp} \gtrsim T_{\rm i,\|}$).
The strongly anisotropic heating at $\beta_{\rm i}\lesssim1$ can be interpreted as the result two dominant perpendicular-heating processes: ion stochastic heating (iSH) and ion cyclotron heating (iCH). 
In fact, iSH is strongly sensitive to the magnetic-fluctuation amplitude at the ion gyro-scale $\delta B_{\rho}/B$~\citep[see discussion in \S4 of][and references therein]{CerriAPJ2021}, and injecting fluctuations with amplitudes $\delta B/B\sim 0.3$ at $k_\perp d_{\rm i}\sim1$ can significantly enhance this mechanism.
On the other hand, injecting fluctuations with non-negligible $k_\|d_{\rm i}$ wavevectors, also produces favourable conditions for iCH to occur~\citep[e.g.,][and references therein]{HollwegIsenberpJGRA2002}.
The less-aniostropic heating in our $\beta_{\rm i}=4$ simulation can instead be explained by the enhancement of both ion-cyclotron heating in the perpendicular direction~\citep[again, because of finte-$k_\|$ effects; ion stochastic heating is instead suppressed at large $\beta_{\rm i}$, see][]{ChandranAPJ2010,CerriAPJ2021} and transit-time damping~\citep[TTD;][]{BarnesPOF1966} in the parallel direction.
In fact, TTD damping is mediated by the magnetic-mirror force and our simulations exhibit an increasingly important magnetic compressibility at sub-ion scales as $\beta_{\rm i}$ increases (Figure~\ref{fig:MagCompr}, left panel).
We also note that, although gyrokinetics operates in a completely different regime than the one investigated here, a larger parallel heating of ions due to TTD of compressive fluctuations is consistent with heating models based on hybrid-gyrokinetic simulations~\citep{KawazuraPRX2020}; and that also ion Landau damping (iLD) of KAW/IKAW fluctuations contributes to parallel ion heating in all HVM simulations, being more important at larger $\beta_{\rm i}$~\citep[e.g., see][]{HowesMNRAS2010,KunzJPP2018,KawazuraPRX2020}.

\section{Conclusions}\label{sec:conclusions}

In this paper, we focused on two main objectives. Firstly, we examined how the occurrence of electron-only reconnection can be triggered in  kinetic-range turbulence. This unique form of reconnection is characterized by an outflow predominantly carried by electrons alone and has been recently observed in various regions in the near-Earth's environment suach as the magnetosheath~\citep{Pha18, StawarzAPJL2019, Sta22}, the magnetopause~\citep{Hua21}, and in the magnetotail~\citep{Wan20}. Secondly, we explored how turbulence conditions that favour the emergence of electron-only reconnection impact  the turbulent heating of ions. 
To that end, we conducted 3D hybrid-Vlasov simulations of sub-ion-scale plasma turbulence with quasi-isotropic, compressible injection close to the ion scales. The fully kinetic ions are coupled to fluid electrons through a generalized Ohm’s law in which electron inertia terms provides the physical mechanism for collisionless magnetic reconnection. 
We have considered three different ion beta values, $\beta_{\rm i} = 0.25, 1, 4$, and ion-to-electron temperature ratios $\tau=T_{\rm 0i}/T_{\rm 0e}=2.5, 10, 40$, respectively. This enable use to probe regimes where $\rho_{\rm i} < d_{\rm i}$, $ \rho_{\rm i} = d_{\rm i}$, and $ \rho_{\rm i} > d_{\rm i}$, respectively, and further disentangle the role of the ion Larmor radius scale $\rho_{\rm i}$ with respect to the ion inertial length $d_{\rm i}$. Moreover, the values of $\tau>1$ are such that $d_{\rm e}>\rho_{\rm e}$ always holds.
These simulations are initialized by injecting fluctuations with $\delta B_{\rm rms}/B_0\sim0.5$ close the ion inertial scale, within the range $0.33 \lesssim k_{\rm inj} d_{\rm i} \lesssim 1$ (i.e., $19\lesssim\lambda_{\rm inj}/d_{\rm i}\lesssim 6$). Our work follows on that of \citet{CalifanoFRP2020} in 2.5D and the initialization reflects conditions similar to those found in the magnetosheath, where ions are typically hotter than electrons, $\tau>1$, and the correlation length of the turbulence exhibiting amplitudes $\delta B/B\sim0.5$--$1$ is estimated to be less than $20 d_{\rm{i}}$~\citep{ChenBoldyrevAPJ2017,Sta22}.

When turbulence is fully developed, the $B_\perp$ spectrum exhibits a $\propto k_\perp^{-8/3}$ scaling at scales $d_{\rm i}^{-1} \lesssim k_\perp < d_{\rm e}^{-1}$, as predicted by intermittency corrected kinetic-Alfv\'en wave (KAW) turbulence~\citep{BoldyrevPerezAPJL2012} and also observed in 3D kinetic simulations~\citep[e.g.,][and references therein]{CerriFSPAS2019}. 
At smaller scales, starting slightly above $k_\perp d_{\rm e}\approx 1$, the $B_\perp$ spectrum steepens to $k_\perp^{-11/3}$, compatible with inertial kinetic-Alfv\'en wave (IKAW) and inertial whistler wave (IWW) turbulence~\citep{ChenBoldyrevAPJ2017}. These spectral slopes are found to be consistent over the range of $\beta_{\rm i}$ values that we have investigated.
Analysis of the scale-dependent magnetic compressibility, $\delta B_\parallel^2 / \delta B_{\perp}^2$, indeed reveals a transition from KAW to IKAW and, subsequently, to IWW type of fluctuations as the energy cascades across $k_\perp d_{\rm e} \sim1$. In the KAW/IKAW regime, density fluctuations are significant, while in whistler turbulence (and EMHD regime), they become negligible. Our simulations show a drop in the density spectrum at sub-ion scales particularly pronounced for $\beta_{\rm i}=4$. This suggests a more negligible ions' response and a more effective transition into the EMHD regime at sub-ion scales for larger $\beta_{\rm i}$, which in turn represents the premise to efficiently develop electron-only reconnection events. 
The velocity spectra show a noticeable decoupling between ions and electrons, which is indeed more prominent as $\beta_{\rm i}$ increases, with less energy being transferred to the small scales of both parallel and perpendicular ion velocities. This decoupling arises due to ions becoming less mobile, thus inhibiting the formation of ion outflows during reconnection events, while electron outflows persist within an electron dissipation region (EDR). 

In our setup, ion heating turns out to be significantly anisotropic at $\beta_{\rm i}\leq 1$, with ions preferentially gaining thermal energy perpendicular to the magnetic-field direction (i.e., $T_{\rm i,\perp} > T_{\rm i,\|}$), while it becomes only marginally anisotropic for $\beta_{\rm i}=4$ (i.e., $T_{\rm i,\perp} \gtrsim T_{\rm i,\|}$). This is the result of ion stochastic heating (iSH) and ion cyclotron heating (iCH) both being the dominant heating processes at $\beta_{\rm i}\lesssim1$~\citep[these mechanisms induce only perpendicular heating, while ion Landau damping (iLD) that would provide parallel heating is a subdominant process in low-$\beta_{\rm i}$ turbulence; see, e.g.,][]{ArzamasskiyAPJ2019,CerriAPJ2021}. At $\beta_{\rm i}=4$, instead, a comparable role of iSH/iCH and transtit-time damping (TTD; which is driven by the enhanced magnetic compressibility, $\delta B_\|^2/\delta B_\perp^2$, of sub-ion-scale fluctuations at larger $\beta_{\rm i}$) provide significant heating both perpendicular and parallel to the magnetic-field direction, respectively.
This $\beta_{\rm i}$-dependent anisotropic heating is found to be also consistent with the ion-cyclotron instability (iCI) possibly playing a relevant role in confining the plasma temperature anisotropy $T_{\rm i,\perp}/T_{\rm i,\parallel}$ within values that belong to a marginally stable range for that instability.

Our results also show that a larger fraction of cascading magnetic energy converts into ion heating as $\beta_{\rm i}$ increases.  Specifically, at $\beta_{\rm i}=0.25$, $1$, and $4$, we observe $Q_{\rm i}/\varepsilon\approx 69\%$, $91\%$, and $96\%$, respectively.
For $\beta_{\rm i}=0.25$, the ion heating estimated from our simulation is consistent with previous results obtained by hybrid-PIC simulations of anisotropic Alfv\'enic turbulence at  $\beta_{\rm{i}}<1$~\citep{ArzamasskiyAPJ2019,CerriAPJ2021} as well as with full-PIC simulations of large-amplitude, compressive turbulence with small-scale injection~\citep{RoytershteynRSPTA2015,Roy22}. However, for $\beta_{\rm i}\geq 1$, we observe an enhanced ion heating with respect to previous simulations of Alfv\'enic turbulence in the same $\beta_{\rm i}$ regimes~\citep{ArzamasskiyAPJ2019,Arz23}. This emphasize the strong sensitivity of ion heating in collisionless space and astrophysical plasmas to the separation between injection scales $\lambda_{\rm inj}$ and the ion gyroradius $\rho_{\rm i}$, with a possibly weaker dependence on the separation between $\lambda_{\rm inj}$ and $d_{\rm i}$ (i.e., while our simulations all have a fixed separation of scales in terms of $d_{\rm i}$ that is smaller than the one employed in hybrid-PIC simulations, the scale separation with respect to $\rho_{\rm i}$ in our $\beta_{\rm i}=0.25$ simulation is similar to the one employed in these previous simulations and we indeed find an ion heating compatible with them). This separation of scales, as well as finite-$k_\|$ effects, can thus strongly affect how the ion-to-electron heating ratio, $Q_{\rm i}/Q_{\rm e}$, depends on the ion plasma beta, $\beta_{\rm i}$~\citep[see, e.g., discussion in][]{Howes_arXiv2024}. Interestingly, our ion-to-electron heating ratio for $\beta_{\rm i}=4$ is in good agreement with the $Q_{\rm i}/Q_{\rm e}$ inferred from MMS measurements within the Earth's magnetosheath in the same range of ion plasma beta~\citep{Roy22}, further validating the applicability of our simulation results to such environment and, in general, to the high-$\beta_{\rm i}$ regime.

Overall, the present study provides new insights for understanding how electron-only reconnection can emerge in fully turbulent, collisionless plasmas such as the Earth's magnetosheath and the bow shock transition region, where small scale current sheets decoupled from the ion dynamics have been discovered (and are now routinely observed) by MMS.
In particular, we have disentangled the role of ion micro-scales, showing that the relevant parameter to trigger electron-only reconnection is the separation between the injection scales and the ion gyroradius $\rho_{\rm i}$ (rather than the ion inertial length $d_{\rm i}$).
Moreover, the results regarding ion turbulent heating also have implications for other collisionless astrophysical environments where various plasma processes, like micro-instabilities or shocks, could inject energy directly near the ion-kinetic scales. 
This is likely the case, for instance, in high-$\beta$ plasmas such as those found in the intracluster medium (ICM) of galaxy clusters.

\begin{acknowledgements}
The authors warmly acknowledge constructive comments from the anonymous referee that helped to improve the quality of the manuscript. We also kindly acknowledge comments and references received from R.~Bandyopadhyay, following our posting of the original manuscript to arXiv. Computations were performed on the HPC system Raven at the Max Planck Computing and Data Facility.
S.S.C. is supported by the French government, through the UCA$^\text{JEDI}$ Investments in the Future project managed by the National Research Agency (ANR) with the reference number ANR-15-IDEX-01, and by the ANR grant ``MiCRO'' with the reference number ANR-23-CE31-0016.
\end{acknowledgements}

\bibliographystyle{apalike}
\bibliography{biblio}

\end{document}